\documentclass[11pt,a4paper]{article}
\usepackage{epsfig,axodraw}
\usepackage{array,cite,amsmath,amssymb,multirow}
\usepackage[flushleft]{threeparttable}
\usepackage{subcaption}
\usepackage{kantlipsum}
\usepackage[usenames,dvipsnames]{xcolor}
\usepackage[breakable, theorems, skins]{tcolorbox}
\tcbset{enhanced}

\usepackage{lineno}
\usepackage{xspace}
\usepackage{listings}
\usepackage{hhline}
\lstdefinestyle{mystyle}{
    backgroundcolor=\color{lightgray}
}
\newcommand{\newc}{\newcommand}
\newcolumntype{M}[1]{>{\centering\arraybackslash}m{#1}}
\newcolumntype{N}{@{}m{0pt}@{}}
\newc{\rpv}{$\mathrm{\not\!R_p}$}
\newc{\rp}{$\mathrm{R_p}$}
\newc{\real}{\mathcal{R}e}
\newc{\alsm}{{\displaystyle \sum_{\alpha=1,2}}}
\newc{\besm}{{\displaystyle \sum_{\beta=1,2}}}
\newc{\al}{\alpha}
\newc{\sgn}{\mr{sgn}\,}
\newc{\be}{\beta}
\newc{\ga}{\gamma}
\newc{\de}{\delta}
\newc{\sla}{\!\!\!\!\!\not\:\:\!}
\newc{\slab}{\!\!\!\!\!\not\,\,\,}
\newc{\slac}{\!\!\!\!\!\!\!\not\,\,\,\,}
\newc{\met}{$\not\!\!E_T$}
\newc{\cw}{\cos\theta_W}
\newc{\sw}{\sin\theta_W}
\newc{\ssw}{\sin^2\theta_W}
\newc{\ccw}{\cos^2\theta_W}
\newc{\cbe}{\cos\beta}
\newc{\sbe}{\sin\beta}
\newc{\ort}{\frac1{\sqrt{2}}}
\newc{\sh}{\hat{s}}
\newc{\uh}{\hat{u}}
\newc{\tha}{\hat{t}}
\newc{\sa}{\sin\al}
\newc{\ca}{\cos\al}
\newc{\mz}{M_{\mr{Z}}}
\newc{\mw}{M_{\mr{W}}}
\newc{\bv}{$\mathrm{\not\!B}$}
\newc{\lv}{$\mathrm{\not\!L}$}
\newc{\ie}{{\it i.e.\/}\ }
\newc{\lam}{\lambda}
\newc{\cht}{\tilde{\chi}}
\newc{\glt}{\tilde{g}}
\newc{\upt}{\tilde{u}}
\newc{\qkt}{\tilde{q}}
\newc{\qt}{\tilde{q}}
\newc{\elt}{\tilde{\ell}}
\newc{\hgt}{\tilde{H}}
\newc{\nut}{\tilde{\nu}}
\newc{\dnt}{\tilde{d}}
\newc{\ftl}{\mr{\tilde{f}}}
\newc{\psb}{\bar{\psi}}
\newc{\rtt}{2^{1/2}}
\newc{\mut}{\tilde{\mu}}
\newc{\mr}{\mathrm}
\newc{\bath}{\bar{\theta}}
\newc{\tht}{\theta}
\newc{\JC}{{\bf J}}
\newc{\lra}{\longrightarrow}
\newc{\eg}{{\it e.g.\  }}
\newc{\me}{\mathcal{M}\xspace}
\newc{\Mmath}{\mathcal{M}\xspace}
\newc{\mathM}{\mathcal{M}\xspace}
\newc{\dbm}{\partial_\mu}
\newc{\dbmu}{\stackrel{\leftrightarrow\  }{\partial^\mu}}
\newc{\sgm}{\sigma_\mu}
\newc{\transpose}{\intercal}
\newc{\tp}{\intercal}
\newc{\captionB}[2]{\caption[{#1}]{{\small {#2}}}}
\newc{\ahref}[2]{#2}
\newc{\tev}{\,TeV}
\newc{\mev}{\,MeV}
\newc{\py}{\textsf{PYTHIA~8}\xspace}
\newc{\hw}{\textsf{Herwig~7}\xspace}
\newc{\mg}{\textsf{MadGraph~5}\xspace}
\newc{\amc}{\textsf{aMCatNLO}\xspace}
\newc{\ra}{\rightarrow}
\newc{\mathL}{\mathcal{L}\xspace}
\newc{\Lmath}{\mathcal{L}\xspace}
\newc{\Lag}{\mathcal{L}\xspace}
\newc{\zp}{$Z'$\xspace}
\newc{\hwfullshower}{\textsf{HW(Z'+SM\ PS)}\xspace}
\newc{\hwfull}{\textsf{HW(Z'+SM\ PS)}\xspace}
\newc{\hwonlyzp}{\textsf{HW(only Z' PS)}\xspace}
\newc{\hwonly}{\hwonlyzp}
\newc{\mgzphwshower}{\textsf{MG(Z')+HW}\xspace}
\newc{\mgzp}{\textsf{MG(Z')}\xspace}
\newc{\ttbar}{$t\bar{t}$\xspace}

\usepackage{jheppub} 

\title{Searching for New Physics Inside Jets with the Herwig 7 Generalised Parton Shower}

\author[a]{Taehee Kim}
\author[a]{Joon-Bin Lee}
\author[b]{M.R. Masouminia}
\author[c,d]{Michael H. Seymour}
\author[a]{Un-ki Yang}

\affiliation[a]{Department of Physics and Astronomy, Seoul National University, Seoul, Republic of Korea}
\affiliation[b]{Institute for Particle Physics Phenomenology, Durham University, Durham, UK}
\affiliation[c]{Department of Physics and Astronomy, University of Manchester, Manchester, UK}
\affiliation[d]{Department of Theoretical Physics, CERN, Geneva, Switzerland}

\emailAdd{taehee.k@cern.ch}
\emailAdd{joon.bin.lee@cern.ch}
\emailAdd{mohammad.r.masouminia@durham.ac.uk}
\emailAdd{michael.seymour@manchester.ac.uk}
\emailAdd{ukyang@snu.ac.kr}

\abstract{
This study investigates parton shower evolution incorporating both Standard Model (SM) and beyond-the-Standard-Model (BSM) radiation, focusing on the phenomenology of a massive $Z'$ boson. While traditional approaches typically assume direct $Z'$ production in the hard process, the possibility of $Z'$ production within jets, enabled by subsequent emissions in the parton shower, offers a complementary opportunity to probe new physics through jet substructure and event topology. The newly developed \hw~framework supporting BSM parton showers enables efficient simulation of $Z'$ production in the logarithmically enhanced regime. Using a simple BSM benchmark, the minimal $U(1)_{B{\rm -}L}$ extension of the SM, the interplay between the SM and BSM showers is evaluated to identify kinematic features that distinguish $Z'$-induced jets from conventional signatures. BSM-radiation signatures are contrasted with SM backgrounds such as QCD, top-quark, and Drell-Yan production, identifying potential discriminants for experimental searches. Experimental sensitivity at the LHC and prospective future colliders is estimated via statistical-significance projections. We find that $Z'$ bosons produced through parton shower radiation populate non-isolated regions inside jets, providing an avenue for new-physics searches overlooked in traditional analyses.
}

\begin{document}
\noindent{\hfill \small CERN-TH-2026-074}

\noindent{\hfill \small MCnet-26-05}

\noindent{\hfill \small IPPP-26-15\\[0.1in]}
\maketitle
\flushbottom


\section{Introduction}
\label{sec:intro}

Despite its precision and predictive power, confirmed by countless experiments, the Standard Model (SM) is known to be incomplete~\cite{sm1,sm2,sm3,sm4,sm5}. It neither incorporates gravity nor explains dark matter and dark energy, the hierarchy problem, or the observed matter-antimatter asymmetry. Moreover, the SM does not accommodate the small but non-zero neutrino masses~\cite{nu_oscillation_theory, nu_oscillation_kamiokande, nu_oscillation_SNO}, persistent tensions in $B$-meson decay observables~\cite{b_anomaly_summary}, and the long-standing discrepancy in the muon anomalous magnetic moment~\cite{muon_g-2_experiment}, all of which motivate physics beyond the SM (BSM). Meanwhile, state-of-the-art hadron colliders such as the Large Hadron Collider (LHC) and future high-energy machines provide a unique environment to probe these open questions by accessing energy scales where new physics may manifest. A broad programme of TeV-scale searches has therefore been pursued, including mono-lepton~\cite{ATLAS-monolepton}, di-lepton~\cite{EXO-19-019, ATLAS-high_mass-dilepton}, multi-lepton~\cite{ATLAS_multilepton,SUS-16-003,EXO-19-002}, missing-transverse-energy plus jets~\cite{EXO-20-004}, di-jet~\cite{ATLAS-energing_jets-hidden_valley, EXO-22-026}, and multi-jet signatures~\cite{EXO-13-001}. To date, there is no direct evidence from these searches using standard objects. As a consequence, increasing attention has shifted to signals concealed in less explored regions of phase space, including jet-associated soft objects, long-lived particles (LLPs), and hidden-valley, dark-shower, and semi-visible-jet signatures~\cite{HIG-16-017,MA_HIG-18-011,EXO-23-002,EXO-23-010,EXO-19-020}. In parallel, the use of B-parking and scouting data has emerged as a game changer, enabling searches at low mass scales and with very soft leptons~\cite{cms-scouting-parking-run1,cms-scouting-parking-run2,EXO-19-018,EXO-21-005,HIG-19-007,EXO-20-014}. These developments highlight both the continued discovery potential of the LHC in low-scale dynamics and the utility of non-standard reconstruction strategies.

Motivated by this landscape, we focus on an experimentally under-investigated regime, scenarios in which a neutral BSM particle is produced within a jet and subsequently decays into a di-muon final state, leading to non-isolated lepton signatures. Such non-isolated leptons have been studied predominantly as fake leptons, or as signal candidates only in limited regions of phase space~\cite{pheno-BtoMuMu,pheno-BtoMuMu-2,EXO-20-014}.
Historically, this regime was de-emphasised due to substantial QCD-induced backgrounds, and searches were mostly restricted to $m_{\mu\mu}$ below the $B$-hadron mass scale, or to displaced-vertex analyses that are effectively background free. However, the LHC is a QCD factory, and this will be even more pronounced at the HL-LHC~\cite{HL-LHC_physics,HL-LHC_prelim_design,HL-LHC_TDR,HL-LHC_upgrade}, so with improved control of QCD backgrounds, we study non-isolated muons originating from the primary vertex in higher-mass regions.

Given the exploratory nature of this analysis, we adopt a model-agnostic approach, with the aim of characterising how new physics may manifest through non-isolated di-muons and identifying robust kinematic features. The mass range considered lies well above the $B$-hadron scale and the hadronisation scale, whilst remaining within the jet environment, and naturally points to scenarios in which new effects enter through parton shower evolution rather than conventional hard-scattering production. For a first study, we therefore use a minimal and widely adopted benchmark, a $U(1)_{B-L}$ extension of the SM featuring a light $Z'$ boson~\cite{bl4-ufo-1,bl4-ufo-2,bl4-ufo-3}.

In hadron collisions, colour-charged partons generate complex final states due to the non-Abelian structure of QCD. Logarithmically enhanced soft and collinear emissions motivate parton shower algorithms, which resum large logarithms and evolve partons from the hard scale down to hadronisation. Modern generators, including \textsf{Herwig~7}~\cite{Bellm:2025pcw, herwig_manual, herwig7.0_release_note, herwig7.2_release_note, Bewick:2023tfi, Masouminia:2023zhb}, \textsf{PYTHIA~8}~\cite{pythia_manual}, and \textsf{Sherpa}~\cite{Sherpa1.1_manual, Sherpa2.2_manual, Sherpa_BSM}, provide state-of-the-art implementations of this perturbative evolution. Historically, there have been only limited general frameworks for the simulation of BSM radiation in parton showers. While dark-sector showers have been explored extensively, significant theoretical and modelling uncertainties remain~\cite{py-hidden_valley-1,py-hidden_valley-2,snowmass-2021}. Recently, more advanced generalised shower capabilities were introduced into the default angular-ordered shower of \hw~\cite{herwig_bsm_ps,hw-hidden_valley} that supports generic BSM radiation based on UFO model inputs~\cite{UFO_model_file} and provides a spin-unaveraged, model-independent formulation covering all permitted splittings within and beyond the SM~\cite{herwig_bsm_ps}.

The present work utilises the \hw BSM parton shower framework to study $Z'$ radiation in the logarithmically enhanced regime within the $U(1)_{B-L}$ model~\cite{bl4-ufo-1, bl4-ufo-2, bl4-ufo-3}. We compare the kinematics of $Z'$ bosons produced via shower radiation with samples in which the $Z'$ is produced at the matrix-element (ME) level. To simplify the analysis, the $Z'$ boson is assumed to decay exclusively to $\mu^+\mu^-$, providing the cleanest signature at ATLAS and CMS even in jet-rich topologies. Signal samples follow the methodology of Ref.~\cite{herwig_bsm_ps}. Hard di-jet events are generated with \mg, and \hw is used to simulate both SM and BSM showers, then compared with $Z'+2$-jet samples generated in \mg. Relative to Ref.~\cite{herwig_bsm_ps}, the present analysis activates both SM and BSM showers simultaneously, enabling a phenomenological validation of the algorithm in the fully showered regime, and quantifying the impact of QCD logarithmic enhancements on soft and collinear $Z'$ production. Since $Z'$ shower events are inherently accompanied by jets, the dominant backgrounds are di-muons plus jets, notably QCD and top-quark-induced processes, with DY becoming relevant as $m_{Z'}$ approaches $m_Z$. We focus on $5 < m_{\mu\mu} < 50~\mathrm{GeV}$, where contributions from $B$-hadron cascade decays and on-shell $Z$ production are suppressed. Using the standard CMS trigger strategy requiring at least one high-$p_T$ muon, we estimate the signal significance for $Z'$ radiation during shower evolution and derive corresponding 95\% confidence-level bounds, then extend the analysis by incorporating scouting triggers, which are particularly effective for probing $\mathcal{O}(10~\mathrm{GeV})$ mass scales.

This paper is organised as follows:
Section~\ref{sec:bsm shower intro} summarises the ingredients required to simulate BSM radiation together with SM showers in \hw, following Ref.~\cite{herwig_bsm_ps}, with emphasis on the implementation and kinematics of $Z'$ emission in the angular-ordered formalism, and on the assumptions entering the logarithmically enhanced, soft and collinear, regime relevant for in-jet production.
Section~\ref{sec:simulation} describes the event-generation strategy, including the construction of shower-induced $Z'$ samples from di-jet hard events, the comparison to ME $Z'+2$-jet samples, and the definition of the corresponding background samples and baseline object selections used throughout.
Section~\ref{sec:kinematics} studies the characteristic kinematics of collinear $Z'$ radiation, including the mapping between shower variables and reconstructed observables, the resulting jet, di-muon, and jet-muon correlations, and the motivation of selection strategies tailored to non-isolated di-muons embedded in jets.
Section~\ref{sec:result} presents sensitivity projections for representative $Z'$ scenarios, including the impact of trigger requirements, and derives expected exclusion bounds from a statistical treatment of signal and background yields.
Section~\ref{sec:summary} concludes and outlines implications for extending existing LHC searches towards non-isolated di-muon final states produced via parton shower radiation.


\section{Revisiting Angular-Ordered Parton Shower and \boldmath$Z'$ Radiation}
\label{sec:bsm shower intro}

In this section, we introduce basic concepts of \hw's parton shower framework and indispensable ingredients for simulating BSM radiation.
Parton shower algorithms play a central role in the simulation of high-energy collisions based on a perturbative approximation to QCD radiation.
For example, in the collinear limit, the differential cross section of a three-parton final state process can be calculated as
\begin{equation}\label{eq:factorisation}
    d \sigma_{AB\ra abc} \approx d\sigma_{AB\ra ab}\,\frac{\alpha_s}{2\pi} \sum_{i=a,b} \frac{d\qt^2}{\qt^2}\,dz\, P_{i\ra ic}(z,\qt),
\end{equation}
where $d\sigma_{AB\ra ab}$, $\qt$, $z$, and $P_{i\ra ic}(z,\qt)$ are the differential cross section for the underlying $2\to2$ process $AB \rightarrow ab$, the shower evolution variable defined below, the light-cone momentum fraction of the emitted parton ($c$) with respect to the parent parton ($a$ or $b$), and the splitting function, respectively.
For the simplest case such as $q\to qg$ splitting, the splitting function is given by $P_{q\ra qg}(z) = C_F (1+z^2)/(1-z)$.
It is then clear that the collinear divergence is governed by the $1/\qt^2$ term while the soft divergence is governed by the splitting function, $P_{i\ra ic}(z,\qt)$.
Here we assume the quarks and the gluon participating in the process are massless so the splitting function depends only on the light-cone momentum fraction, but the $\qt$ dependence should be reinstated when mass effects are included.

A fundamental variable in a parton shower is the light-cone momentum fraction, \( z \), which quantifies the fraction of the progenitor’s momentum that is carried by one of the daughter partons after a splitting.
Equally crucial is the evolution scale, \( \qt \), defined in a manner that preserves the dot product of the four-momenta of the daughter partons during the branching process, given as~\cite{log_accuracy-hw}:
\begin{equation}
    \qt^2 = \frac{2q_1\cdot q_2 - m_0^2 + m_1^2 + m_2^2}{z(1-z)}.
\end{equation}
It can be interpreted as a measure of the angular separation of two daughter partons multiplied by the energy of the progenitor, i.e. $E_i \theta_{ij}$, in the massless limit~\cite{herwig_manual}.
Angular ordering is naturally achieved by requiring starting evolution scales of $z\qt$ and $(1-z)\qt$ for each daughter parton, respectively.
Unlike transverse-momentum- or virtuality-based showers used in other generators such as \py, the angular-ordered splitting approximation preserves soft-gluon coherence, a vital physical requirement derived from the non-Abelian nature of QCD, without the need for special treatment such as dipole algorithms~\cite{intro_to_AO,intro_to_AO-2}.
There are several choices to impose angular ordering, but the dot-product-preserving scheme is often considered the most natural option, yielding a fully physical solution with a simple condition, preventing unphysically large virtualities, and describing mass effects accurately at leading-logarithmic order.
In the case of $Z'$ radiation, this choice is important to properly handle massive particle emission, where the virtuality can directly affect the final observables and mass effects are significant.

Going back to Eq.~\eqref{eq:factorisation}, the latter part of the equation can be re-interpreted as the differential probability of emitting an additional parton, i.e.
\begin{equation}
    dP_i (\qt) = \frac{\alpha_s}{2\pi} \sum_{j,k} \frac{d\qt^2}{\qt^2} \int_{\qt_0/\qt}^{1-\qt_0/\qt} dz\, P_{i\ra jk}(z,\qt),
\end{equation}
at a given evolution scale, $\qt$. 
Note the integration range is given schematically for intuition, with $\qt_0$ acting as an infrared regulator, and the summation now includes all allowed branchings of the progenitor parton $i$.
It is trivial that we should obtain unity when summing the probability of emitting an additional parton and the probability that the parton $i$ does not radiate.
Therefore, the no-emission probability from a given evolution scale $\qt_0$ down to $\qt_1$, the Sudakov form factor, is written as
\begin{equation}
    \Delta_i(\qt_0^2,\qt_1^2) = \prod_{\qt \in [\qt_0,\qt_1]} \left(1-dP_i\right)
    = \exp \left[ -\frac{\alpha_s}{2\pi} \sum_{j,k} \int_{\qt_0}^{\qt_1} \frac{d\qt^2}{\qt^2} \int_{\qt_0/\qt}^{1-\qt_0/\qt} dz\, P_{i\ra jk}(z,\qt) \right].
\end{equation}

In SM splittings, for cases where the progenitor preserves its flavour after emission, there is at least single-pole behaviour such as $P_{i\ra ij} \sim 1/(1-z)$, resulting in
\begin{equation}
    \Delta_i(\qt_0^2, \qt_1^2) \sim \exp \left( -\frac{\alpha_s}{2\pi} \log^2 \frac{\qt_1^2}{\qt_0^2} \right).
\end{equation}
This exponentiation corresponds to a resummation of all orders in $\left(\alpha_s \log^2(\qt_1^2/\qt_0^2)\right)^n$~\cite{general_purpose_event_generators}.
The parton shower algorithm generates emissions by sampling the corresponding Sudakov form factors, thereby incorporating the dominant virtual corrections associated with unresolved radiation.
As a result, the resummation of these double-logarithmic contributions yields a parton shower with leading-logarithmic accuracy.

A further refinement is the treatment of spin correlations in the radiation processes.
Historically, spin correlations are considered at production and decay processes of massive particles such as top quarks, b-quarks, and tau leptons.
However, it has been highlighted that spin correlations are crucial in parton showers to accurately describe angular distributions~\cite{herwig_manual,spin_correlation-collins,spin_correlation-richardson,herwig7.0_release_note}.
Their impact is more critical in parton showers involving massive EW bosons, the Higgs boson, and even BSM parton showers than in QCD showers commonly assumed to be nearly massless.
In \hw, spin information is thus preserved throughout the event by tracking helicity amplitudes along the evolution of polarised particles.
The spin density matrix formalism is used to propagate spin information through the shower and decay chain.
For example, in the production of a polarised heavy fermion, a spin density matrix is constructed at the production vertex and used to weight the decay amplitudes appropriately.
Such a coherent treatment of spin and colour in showers and decays allows \hw to attain accuracy in differential distributions involving angular observables.
This allows a reliable evaluation of the fraction of muons that are genuinely located inside jets.

Employing these vital ingredients described above, the parton shower algorithm can be implemented using the Monte Carlo (MC) method.
To briefly summarise this procedure, one begins by determining the scale $\qt_1$ at which the next branching occurs, starting from an initial parton at scale $\qt_0$.
Using the MC method, a random number $\rho \in [0,1]$ is selected, and $\qt_1$ is determined such that $\Delta_i(\qt_0^2,\qt_1^2) = \rho$.
If $\qt_1$ is larger than a certain scale $\tilde{Q}$, which marks the transition to non-perturbative physics from the parton shower, a new branching is generated at $\qt_1$.
Conversely, if $\qt_1$ is below $\tilde{Q}$, the evolution of that parton branch is terminated.
The branching at $\qt_1$ is determined using the integrated emission probabilities.
Specifically, the quantity $\mathcal{P}_{i \rightarrow jk}$, defined as
\begin{equation}
    \mathcal{P}_{i \rightarrow jk} = \frac{\int_{z_{\text{min}}}^{z_{\text{max}}} P_{i \rightarrow jk}(z) \, dz}{\int_{z_{\text{min}}}^{z_{\text{max}}} \sum_{j,k} P_{i \rightarrow jk}(z) \, dz},
\end{equation}
is compared against a random number $\lambda \in [0,1]$.
For instance in the case of a gluon line, if $\lambda < \mathcal{P}(g \rightarrow q\bar{q})$, the splitting $g \rightarrow q\bar{q}$ is chosen, while if $\mathcal{P}(g \rightarrow q\bar{q}) < \lambda < \mathcal{P}(g \rightarrow q\bar{q}) + \mathcal{P}(g \rightarrow gg)$ (which equals unity if one considers QCD-only branchings), the splitting $g \rightarrow gg$ is selected.
Through this procedure, the type of branching is determined.
Once the branching type is fixed, another random number $r \in [0,1]$ is selected, allowing the determination of the momentum fraction $z$ by solving
\begin{equation}
    r \int_{z_{\text{min}}}^{z_{\text{max}}} P(z) \, dz = \int_{z_{\text{min}}}^{z} P(z') \, dz'.
\end{equation}
With the values of $\qt$ and $z$ obtained in this way, all remaining kinematic variables, except for the azimuthal angle $\phi$, which is chosen uniformly, can be fully determined.
Furthermore, modern parton shower event generators employ additional techniques, such as rejection sampling, to improve the efficiency of this procedure.
The simulation proceeds by evolving each colour-charged parton from the hard scale of the process down to a non-perturbative infrared cutoff, below which hadronisation models take over.


One of the major strengths of the parton shower algorithm described above lies in the fact that, although QCD parameters such as $\alpha_s$ and the representative $q\to qg$ process were used as examples, no model-specific assumptions were introduced in its logical development.
As a consequence, this method can be extended to theories beyond QCD in an entirely analogous manner.
Parton shower algorithms for QED~\cite{herwig_qed_radiation}, EW boson~\cite{herwig_ewk_radiation, Darvishi:2021het, Masouminia:2026udz}, and BSM boson~\cite{herwig_bsm_ps} radiation have been implemented in \hw.
This implies that $Z'$ radiation is simulated at the same level and in the same manner as the SM shower emissions.

Specifically, the $f \to f' V$ splitting is the simplest scenario for BSM boson radiation, with the interaction described by a vertex of the form
\begin{equation}
    -i \left( g_L P_L + g_R P_R \right) \gamma^\mu.
\end{equation}
The longitudinal polarisation component, which can lead to numerical divergences in the limit $m_V \to 0$, is handled by the Dawson prescription~\cite{Dawson}.
The resulting splitting probability is given by
\begin{align}
    P_{f \to f'V}(z, \qt) &= (|g_R|^2 \rho_+ + |g_L|^2 \rho_-) \left[ \frac{1 + z^2}{1 - z}(1 + m_{f,t}^2) - \frac{1 + z}{1 - z} m_{f',t}^2 - m_{V,t}^2 \right] \nonumber \\
    &\quad + (|g_R|^2 \rho_- + |g_L|^2 \rho_+) z m_{f,t}^2 - 2 \mathrm{Re}(g_L g_R^*) (\rho_+ + \rho_-) m_{f,t} m_{f',t},
\end{align}
where $m_{i,t} = m_i / (\sqrt{z(1-z)}\,\qt)$ and $\rho_\pm$ encode the helicity populations of the incoming fermion~\cite{herwig_bsm_ps}.
This formulation accommodates broad types of couplings such as arbitrary combinations of left- and right-handed couplings, as well as the presence of flavour-changing neutral currents (FCNCs).
Gauge artefacts are explicitly subtracted, resulting in well-behaved splitting kernels.

For the purpose of this study, the expression is further simplified by adopting the minimal $U(1)_{B-L}$ model, in which the left- and right-handed couplings are identical and FCNCs are absent, so that $f'=f$ and $m_f = m_{f'}$~\cite{bl4-ufo-1,bl4-ufo-2,bl4-ufo-3}.
The fermion-to-vector splitting function reduces to
\begin{equation}
    P_{f \to fV}(z,\qt) = \frac{g^2}{2} \left( \frac{1 + z^2}{1 - z} - 2 m_{f,t}^2 - m_{V,t}^2 \right).
\end{equation}
The remainder of this paper focuses on the kinematic features exhibited by the $Z'$ bosons generated through this splitting kernel.


\section{Sample Generation}
\label{sec:simulation}

The sample production strategy is a critical component of this analysis, as the targeted signatures arise from BSM radiation occurring during parton shower evolution rather than from a conventional hard-scattering process. The BSM shower framework implemented in \hw~\cite{herwig_bsm_ps} has been developed recently and has not, to our knowledge, been employed in previous phenomenological or experimental studies. A careful construction of the signal samples is therefore essential to reliably capture the kinematic characteristics of $Z'$ radiation in this setup.

This section describes the procedure for generating the signal samples, together with dedicated validation and background samples, each serving a complementary role. The validation samples are designed to assess the consistency and robustness of the BSM parton shower algorithm, in particular in the collinear region relevant for logarithmically enhanced radiation. The background samples, namely QCD- and top-induced processes, are constructed to characterise the relevant phase space and to quantify the experimental sensitivity in the presence of large SM backgrounds. Taken together, these samples provide the basis for the phenomenological interpretation and for the robustness of the results presented in the following sections.

\subsection{Signal Sample}\label{subsec:signal}

\mg~provides a robust framework for BSM event generation at ME level, automatically constructing amplitudes from UFO model files, including the corresponding vertices and propagators~\cite{UFO_model_file,MG5_BSM}.
This convenience has led many analyses to compute BSM production in \mg, subsequently adding SM parton showers with general-purpose event generators such as \hw~or \py~to simulate complete events.
However, this study targets BSM particles produced within jets, which are associated with logarithmically enhanced QCD radiation.
In such cases, a purely ME-based approach becomes computationally challenging, as it requires the evaluation of high-multiplicity final states once additional QCD radiation is included.
We therefore employ the newly developed BSM parton shower framework in \hw~\cite{herwig_bsm_ps}.
Specifically, the hard scattering process, namely di-jet production, is generated with \mg, while the subsequent $Z'$ radiation is simulated in \hw as part of the combined SM and BSM parton shower evolution.
Although both SM and BSM particles are evolved together, the contribution from BSM radiation is parametrically suppressed.
We therefore employ the MMHT2014lo68cl LO PDF set~\cite{MMHT14} together with the global \textsc{Herwig}~7.3 tune, which has been calibrated to electroweak precision data and does not incorporate BSM radiation effects in the jet modelling~\cite{Masouminia:2023zhb,Bewick:2023tfi}.
All $Z'$ bosons are assumed to decay into a muon pair for simplicity, since hadronic decays are inseparable from the jet constituents and electrons inside jets are experimentally challenging in this topology.

This analysis does not target a specific model, but instead assumes generic new physics produced within a jet.
We therefore employ a representative benchmark featuring a $Z'$ boson, the minimal $U(1)_{B-L}$ extension of the SM~\cite{bl4-ufo-1, bl4-ufo-2, bl4-ufo-3}.
This choice provides a minimal and widely used reference point, allowing comparison with existing literature and straightforward extensions to more elaborate scenarios.

The collider-relevant interaction is taken to be
\begin{equation}\label{eq:lagrangian}
    \Lag \supset \left( g_{qZ'} \, \Bar{q} \gamma^\mu q + g_{\ell Z'} \, \Bar{\ell} \gamma^\mu \ell \right) Z'_\mu,
\end{equation}
where no mixing with SM gauge bosons is introduced.
The SM Lagrangian remains otherwise unchanged.
Other interactions that may arise in specific ultraviolet completions are not relevant for the collider phenomenology considered here~\cite{bl4-ufo-1,bl4-ufo-2,bl4-ufo-3}.
The purely vector-like nature of the $Z'$ coupling implies identical interactions with left- and right-handed fermions, thereby simplifying the theoretical structure.
Furthermore, by fixing $\mathrm{BR}(Z'\to\mu\mu)=1$ for the purposes of this study, the production phenomenology is controlled by $g_{qZ'}$ and $M_{Z'}$.

When generating signal samples using \hw, special attention must be paid to the differences between resonant particles originating from the parton shower and those produced in the hard process.
As with the hard process, the rate of $Z'$ production through the parton shower increases with the square of the coupling \( g_{qZ'} \).
However, in the parton shower, the rate of events containing a radiated $Z'$ is bounded by the underlying di-jet event rate, whereas at the ME level, the $Z'$ production rate is not subject to such a bound from the di-jet rate.
It also becomes possible for multiple $Z'$ bosons to be emitted within a single event.
Such events are not desirable here, since the intended regime corresponds to sufficiently small BSM couplings that multi-$Z'$ emission is negligible.
To prevent this, the coupling must be carefully tuned\footnote{Even when multiple $Z'$ bosons are not produced, an excessively large coupling can drive $Z'$ emission into regions where the parton shower phase-space construction breaks down.
In this case, parton shower algorithms, which are designed to describe soft and collinear physics, cease to provide a reliable approximation.}.
Furthermore, choosing a coupling that is too small substantially suppresses the probability of $Z'$ radiation in di-jet events, leading to inefficient signal production.

\begin{figure}
    \centering
    \includegraphics[width=0.49\linewidth]{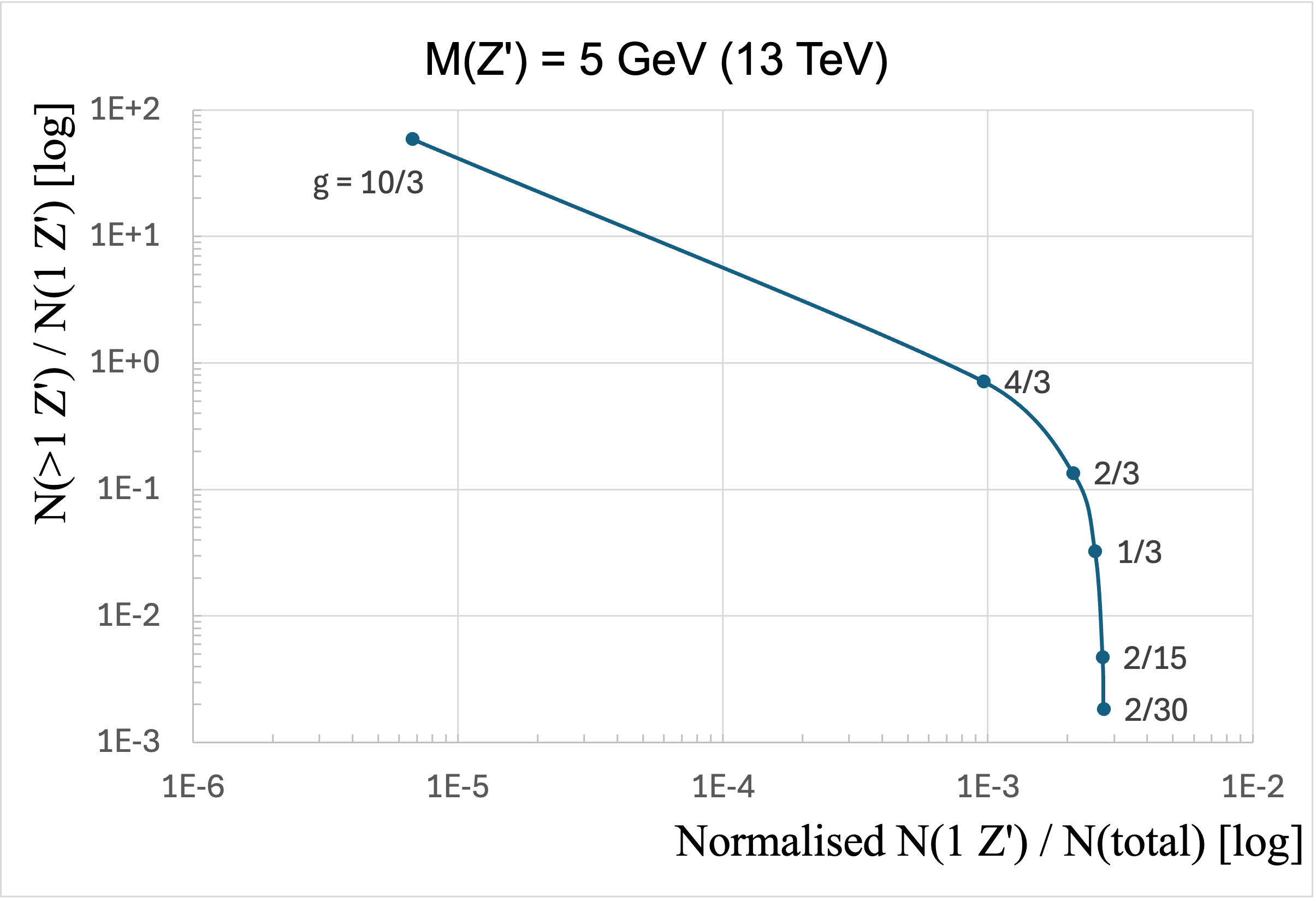}
    \includegraphics[width=0.49\linewidth]{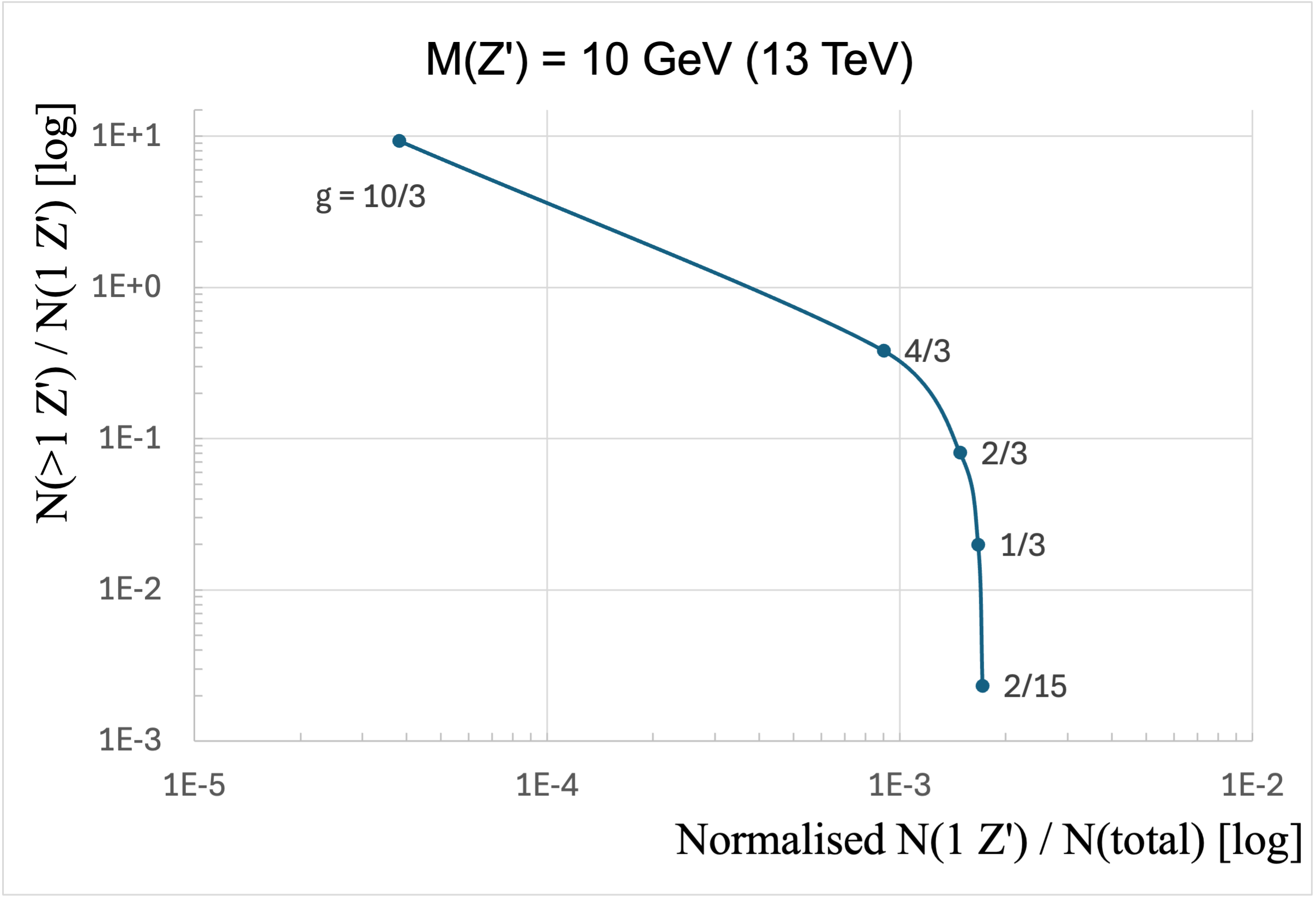}
    \includegraphics[width=0.49\linewidth]{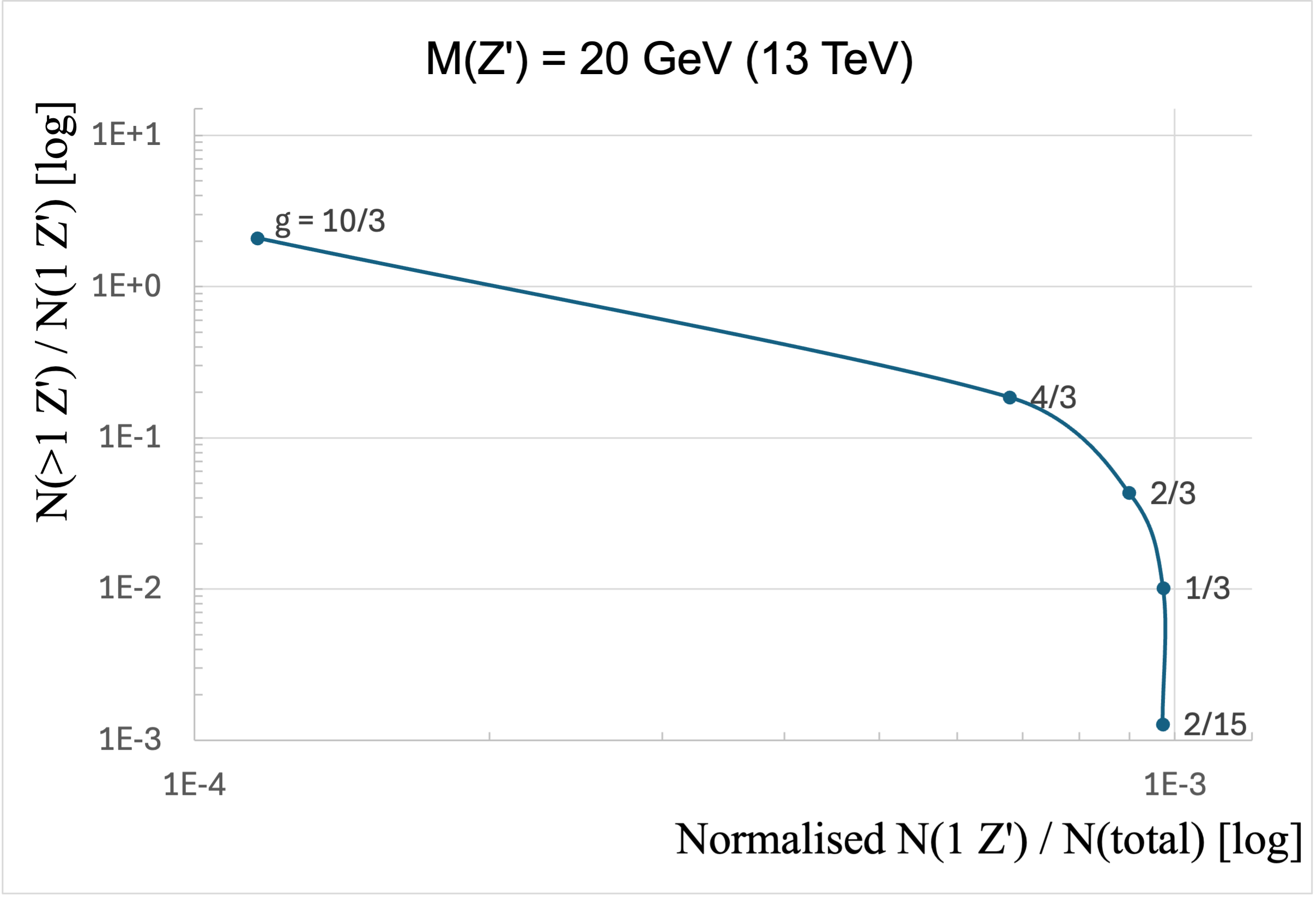}
    \includegraphics[width=0.49\linewidth]{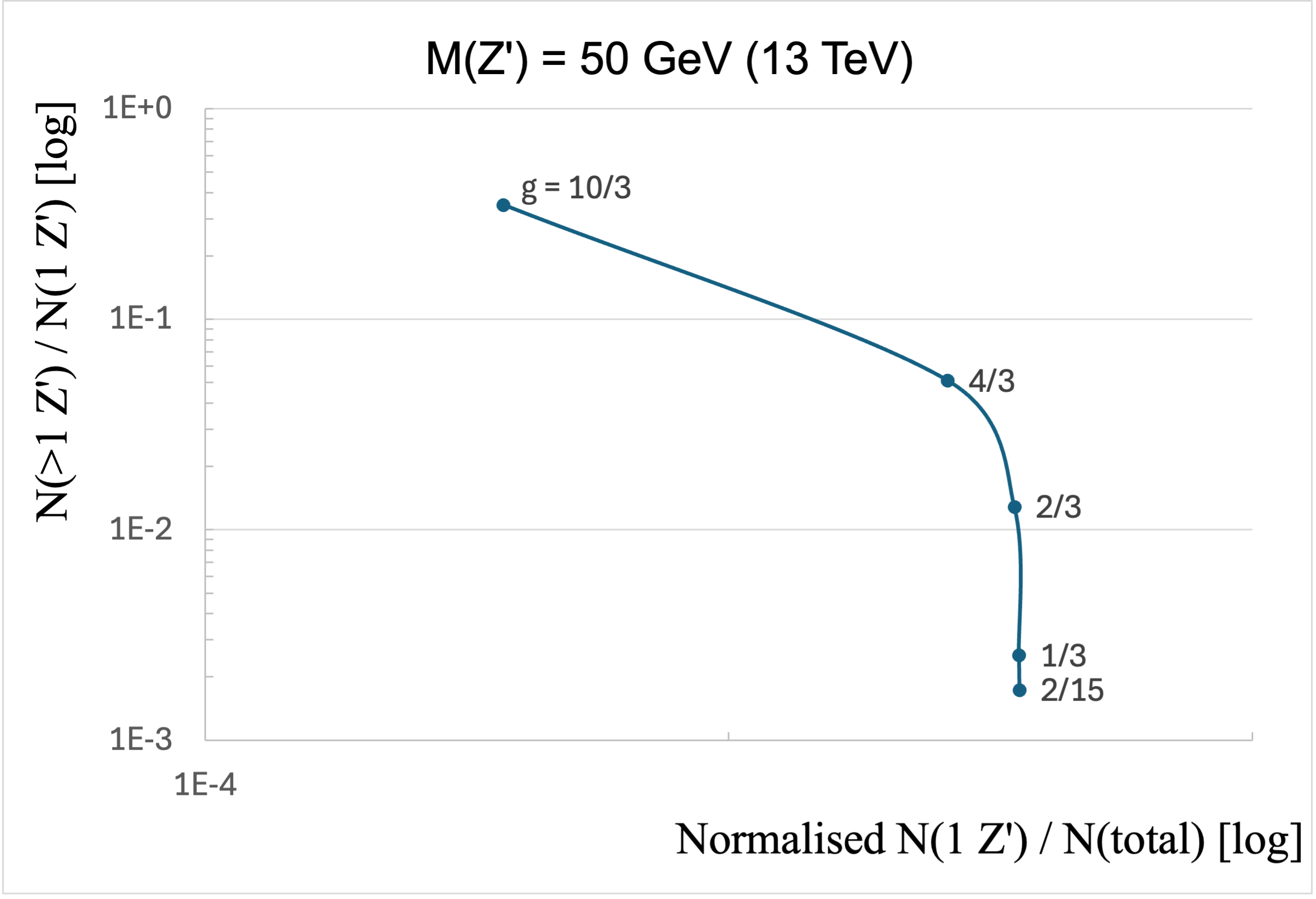}
    \caption{Coupling-dependence study of $Z'$ radiation in the parton shower for $M_{Z'}=5$, 10, 20, and 50~GeV.
    Each panel shows the ratio of events containing more than one $Z'$ boson to those containing exactly one $Z'$ boson, $N(>1~Z')/N(1~Z')$, as a function of the fraction of di-jet events with exactly one $Z'$ boson, $N(1~Z')/N(\text{total})$ normalised by $1/g_{qZ'}^{2}$.
    }
    \label{fig: coupling test}
\end{figure}

To address this issue, test samples are generated for $M_{Z'}=5$, 10, 20, and 50~GeV with several coupling choices ranging from $g_{qZ'}=2/30$, corresponding to the default choice in the original model file~\cite{bl4-ufo-1,bl4-ufo-2,bl4-ufo-3}, up to $10/3$. 
The hard process is di-jet production with a matrix-element level transverse-momentum requirement of $200 < p_T < 300~\mathrm{GeV}$ imposed on the outgoing partons, generated using \mg. Figure~\ref{fig: coupling test} shows $N(>1~Z')/N(1~Z')$ as a function of $N(1~Z')/N(\text{total})$, where $N(1~Z')$ is normalised by $1/g_{qZ'}^2$.
Since events with more than one $Z'$ boson are removed from the signal definition, their contribution must be statistically negligible.
We therefore select coupling values satisfying $N(>1~Z')/N(1~Z') < 0.01$ as working points.
In this regime, $N(1~Z')/N(\text{total})$ normalised by the squared coupling becomes independent of $g_{qZ'}$, indicating that the shower-induced $Z'$ production rate scales as $g_{qZ'}^2$, analogous to the leading coupling dependence of hard-scattering production.

In the following sections, the samples generated by the aforementioned method are denoted as ``\hwfull'', indicating that the \zp~boson is simulated within the full parton shower framework of \hw, including both SM and BSM radiation.
In addition, to reproduce the distributions studied in Ref.~\cite{herwig_bsm_ps} for comparison, the same samples are produced with the SM parton shower switched off, retaining only the $Z'$ parton shower component; these are denoted as ``\hwonly''.
This ensures that the comparison remains meaningful and allows us to isolate the effect of BSM radiation relative to the complete shower evolution.

\subsection{Validation Sample}
\label{sec:validation_sample}

\begin{figure}
    \centering
    \includegraphics[width=0.25\linewidth]{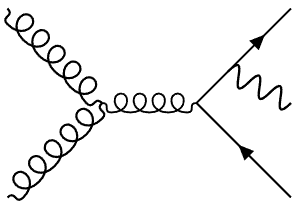} \hspace{0.5cm}
    \includegraphics[width=0.25\linewidth]{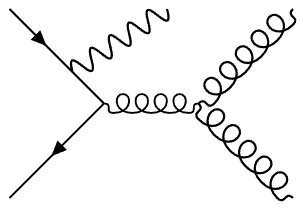} \hspace{0.5cm}
    \includegraphics[width=0.23\linewidth]{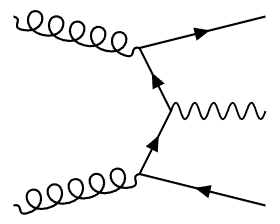}
    \caption{Three typical Feynman diagrams that can produce a \zp boson in proton-proton collisions. The left, centre, and right panels show FSR-like, ISR-like, and quark-quark fusion (qqF) topologies, respectively.}
    \label{fig:feynman diagram}
\end{figure}

The $Z'$ radiation samples generated with the \hw BSM parton shower are validated by comparison with reference samples in which the $Z'$ contribution is computed at ME level using \mg, following the strategy adopted in Ref.~\cite{herwig_bsm_ps}.
The key difference with respect to the earlier study concerns the treatment of the parton shower in \hw.
In the previous analysis, the shower evolution was restricted to a single BSM emission in final-state radiation (FSR), as illustrated in the first panel of Figure~\ref{fig:feynman diagram}, enabling a direct comparison with fixed-order (FO) calculations.
In contrast, the present study allows full shower evolution, including both SM and BSM radiation in initial-state radiation (ISR) as well as FSR.
This provides a more realistic assessment of the shower-induced signatures expected from the complete implementation.

To enable a fair comparison with the full parton shower samples, we extend the validation strategy beyond that used in Ref.~\cite{herwig_bsm_ps}.
While the previous work considered only FSR-like topologies at ME level, we here generate $Z'+2$-jet samples in \mg including all relevant diagrams, encompassing ISR-like topologies as well as quark-quark fusion (qqF) production, as illustrated by the second and third diagrams in Figure~\ref{fig:feynman diagram}.
The resulting events are subsequently passed to \hw to apply the SM parton shower.
In the following, this validation sample, consisting of \mg-generated $Z'+2$-jet events supplemented by the SM shower in \hw, is denoted as ``\mgzphwshower''.

In addition, to facilitate a direct comparison with the \hwonly~samples, we also generate a second set of validation samples without any parton showering.
These samples provide pure FO predictions including all relevant diagrams, and are referred to as ``\mgzp''.
In contrast to Ref.~\cite{herwig_bsm_ps}, where the FO comparison was restricted to the FSR-like contribution, the ``\mgzp'' samples include the full $Z'+2$-jet ME content used for ``\mgzphwshower''.


\subsection{Background Sample}
\label{sec:background_sample}

There are several SM processes that can mimic non-isolated di-muon signals. It is particularly difficult to distinguish these backgrounds, since they may also produce two muons in association with a jet. Representative examples are QCD processes involving heavy-flavour decays, top-quark-induced backgrounds, and Drell-Yan (DY) production accompanied by additional jets (DY+$N$-jet). In particular, QCD contributes through multiple channels and dominates across most of the relevant phase space due to its large cross section.

For instance, a single $B$ hadron can yield two muons via cascade decays, representing the dominant background in the di-muon mass region below 5~GeV, the approximate scale of the $B$-hadron mass. Above this threshold, an important contribution arises from gluon splitting, $g \to b\bar{b}$ or $g \to c\bar{c}$, where each heavy quark can subsequently produce a muon. At higher di-muon masses, backgrounds from top-quark production become increasingly relevant. The DY process can also mimic signal-like events when leptons from $Z/\gamma^*$ become collimated with a jet, although its contribution is suppressed in the mass range considered here. Accurate modelling of these background components is therefore essential for a reliable sensitivity estimate.

The QCD background samples are generated following the standard CMS MC production strategy, using \py at leading order.
Since this analysis employs both the standard muon trigger and a scouting strategy accepting muons at the GeV scale, the QCD background must be simulated down to relatively low transverse momenta. Accordingly, the minimum transverse momentum of the hard process is set to 20~GeV.

At such low scales, the steeply falling QCD spectrum leads events to be concentrated near $p_T \sim 20$~GeV, whereas non-isolated di-muon signatures predominantly arise from higher-$p_T$ configurations, since being contained within a jet implies a small angular separation between the two muons, typically enforced via a maximum $\Delta R$ requirement. To efficiently populate the relevant phase space, the samples are therefore produced in exclusive bins of the hard-scattering scale, defined in terms of the partonic transverse momentum~$\hat{p}_T$. The number of events in each bin is chosen to minimise the total statistical uncertainty of the combined sample while keeping the overall event count as small as possible.
The total statistical uncertainty is defined as
\[
\text{Total statistical uncertainty} = \frac{\sqrt{\sum_i \sigma_i^2 \epsilon_i / N_i}}{\sum_i \sigma_i \epsilon_i},
\]
where $N_i$, $\epsilon_i$, and $\sigma_i$ are the number of generated events, the pre-selection efficiency, and the cross section of each sample, respectively. It is minimised when the number of generated events per bin satisfies $N_i \propto \sigma_i \sqrt{\epsilon_i}$. This binning strategy substantially improves the statistical precision in the high-$p_T$ region, which dominates the background composition after event selection, while retaining sufficient coverage at lower transverse momenta.

Light-hadron decays, e.g. charged pions, charged kaons, and long-lived neutral kaons, are enabled in order to retain a realistic description of muons originating from hadronic activity inside jets. The generated events are subsequently filtered to enhance the fraction of events containing two muons within the detector acceptance, thereby increasing the effective efficiency of the background samples without biasing the underlying QCD dynamics. Parton showering and hadronisation are performed using the CP5 tune, which provides a modern and well-validated description of QCD radiation and underlying-event activity at the LHC~\cite{CP5_tune}. This setup allows the dominant QCD backgrounds to be modelled with sufficient accuracy in the kinematic regions relevant for this analysis, while keeping the computational cost under control.

Top-quark background samples are generated using the \textsf{POWHEG} generator at next-to-leading order (NLO) accuracy for both \ttbar and $tW$ processes~\cite{powheg1,powheg2,powheg3,powheg-tt}.
The hard-scattering events are interfaced with \py for parton showering and hadronisation, employing the CP5 tune to ensure a consistent description of QCD radiation and underlying-event activity.
Spin correlations in top-quark decays are retained through the dedicated decay treatment implemented in \textsf{POWHEG}, which is essential for accurately modelling the kinematic properties of decay products.

For the \ttbar process, separate samples are generated for the di-leptonic, semi-leptonic, and fully hadronic decay channels. Although this analysis targets muon-based signatures, the fully hadronic channel is explicitly included. This is motivated by the fact that muons can still arise from the subsequent decays of $b$ hadrons produced in hadronic top decays, thereby contributing non-isolated muons inside jets that populate the signal-like phase space.


\section{Kinematics and Event Features of \boldmath$Z'$ Radiation}
\label{sec:kinematics}

Since this analysis constitutes the first phenomenological application of the \hw BSM parton shower framework with simultaneous SM and BSM evolution, a detailed understanding of whether the generated signal samples correctly populate the logarithmically enhanced phase space, and of the characteristic signatures they exhibit, is essential.
To this end, following the strategy adopted in earlier studies~\cite{herwig_bsm_ps}, we examine parton-level kinematic distributions using the aforementioned samples, \hwfull, \hwonly, \mgzphwshower, and \mgzp.
This comparison allows us to disentangle differences between $Z'$ bosons produced via parton shower radiation and those generated at fixed order (FO), as well as to isolate the impact of the presence or absence of the SM parton shower on the resulting kinematic features.

One crucial prerequisite for examining these parton-level kinematics is the unambiguous identification of the quark that radiates the $Z'$ boson.
For samples in which the $Z'$ bosons are generated within \hw, the shower history is explicitly available, rendering the identification of the radiating quark straightforward.
By contrast, when the $Z'$ boson is simulated at the ME level using \mg, such an identification is inherently ambiguous.

To address this, we follow the same prescription adopted in the original \hw BSM parton shower study~\cite{herwig_bsm_ps}.
Specifically, we first identify the two partons ($p_1$ and $p_2$) originating from the hard process and determine, for each of them, the transverse momentum of the $Z'$ boson evaluated in a reference frame where the $p_i Z'$ system defines the longitudinal axis.
The parton with respect to which the $Z'$ boson exhibits the smaller transverse momentum is then selected as the partner quark, corresponding to the most collinear radiation configuration.

It is worth noting that, unlike in the earlier study and in the \hwonly samples, where the SM parton shower is switched off, additional QCD radiation, such as gluon emission, may occur prior to the $Z'$ emission in the \hwfull samples.
In such cases, we do not use the partons immediately after the hard process, but instead identify the emitter at the stage following the $Z'$ emission.
By defining the partner quark in this way, we eliminate ambiguities arising from intermediate QCD emissions and ensure a consistent interpretation of the parton-level kinematics.

\begin{figure}
    \centering
    \includegraphics[width=0.49\linewidth]{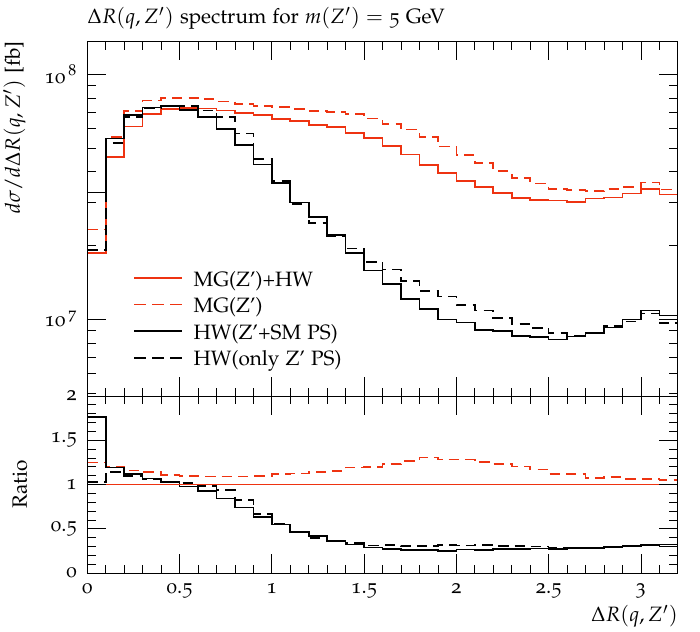}
    \includegraphics[width=0.49\linewidth]{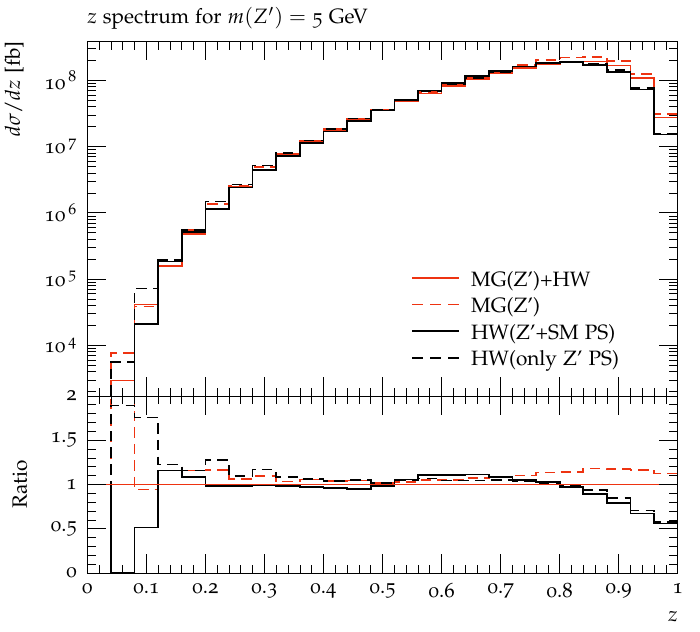}
    \includegraphics[width=0.49\linewidth]{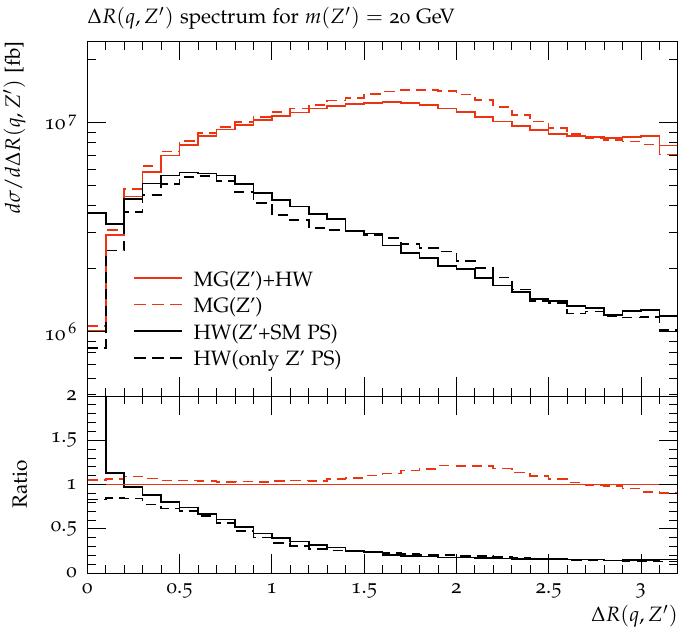}
    \includegraphics[width=0.49\linewidth]{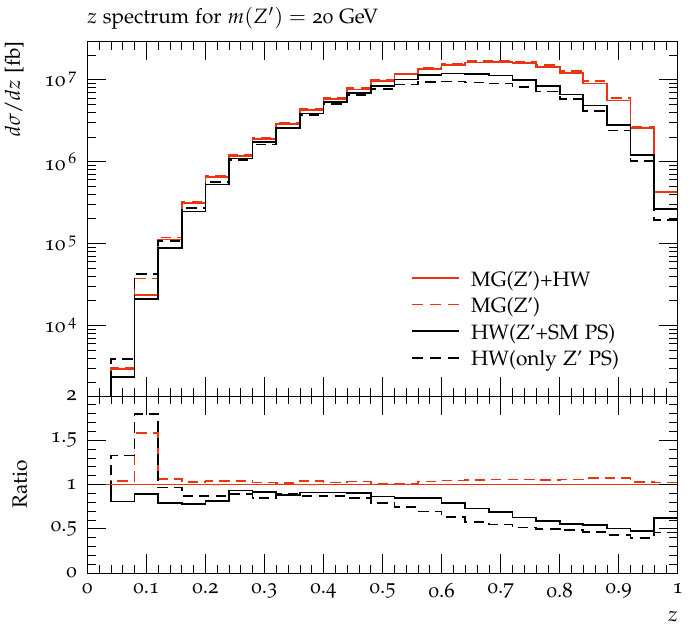}
    \includegraphics[width=0.49\linewidth]{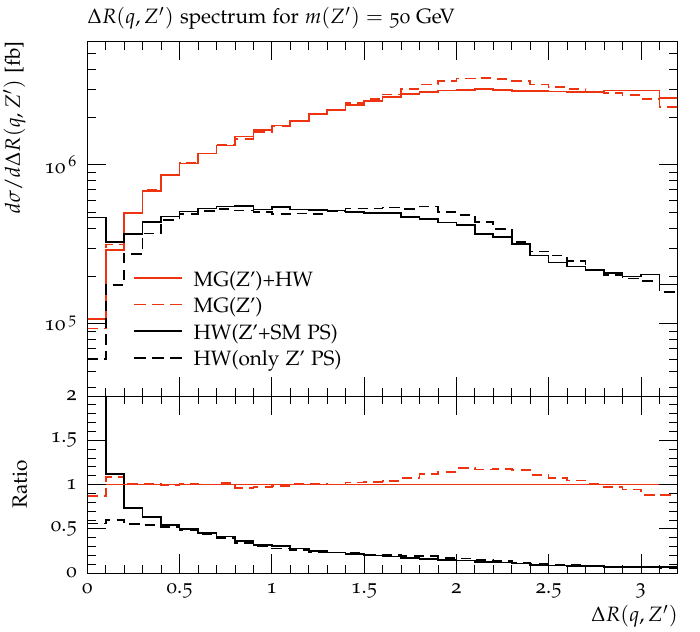}
    \includegraphics[width=0.49\linewidth]{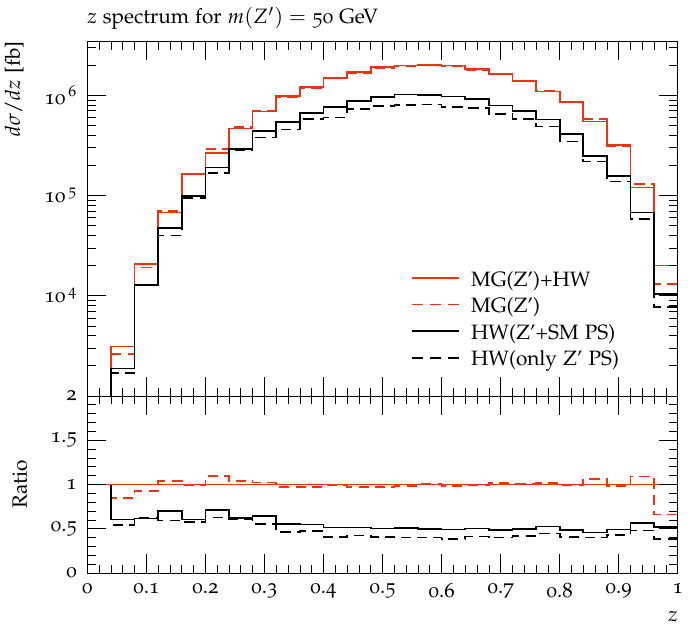}
    \caption{
    Parton-level validation of the $Z'$ radiation mechanism for representative mass points, $M_{Z'} =$ 5, 20 and 50~GeV.
    The distributions show the angular separation $\Delta R(q, Z')$ (left) and the transverse-momentum fraction carried by the quark within the $qZ'$ system, namely $z$ (right).
    Results are shown for the signal and validation samples discussed in Sec.~\ref{sec:simulation}.
    For the $z$ distributions, only events satisfying $\Delta R(q, Z')<1$ are included to suppress ISR-like and qqF topologies and to isolate the logarithmically enhanced phase space relevant for parton shower radiation.
    }
    \label{fig: parton validation}
\end{figure}

Figure~\ref{fig: parton validation} presents the parton-level kinematic distributions between the identified partner quark ($q$) and the $Z'$ boson for typical mass points, $M_{Z'}=5$, 20, and 50~GeV.
Shown are the angular separation $\Delta R(q,Z')$ and the momentum-sharing variable defined as $p_T(q)/(p_T(q)+p_T(Z'))$.
The latter observable corresponds to the energy-fraction variable, commonly denoted as $z$, that appears in the parton shower splitting kernel.
For clarity, Table~\ref{tab:samples_summary} summarises the generator configurations used in these comparisons, following the definitions in Sec.~\ref{sec:simulation}.

\begin{table}[t]
\centering
\begin{tabular}{lccc}
\hline
 & \mg\ (ME) & \multicolumn{2}{c}{\hw\ (shower)} \\
Sample & Hard process & $Z'$ shower & SM shower \\
\hline
\mgzphwshower & $pp\to Z'jj$ & off & on  \\
\mgzp         & $pp\to Z'jj$ & off & off \\
\hwfull       & $pp\to jj$   & on  & on  \\
\hwonly       & $pp\to jj$   & first-emission only  & off \\
\hline
\end{tabular}
\caption{Summary of the event sample configurations. The table specifies the matrix-element process and the treatment of $Z'$ and SM radiation in the \hw shower.}
\label{tab:samples_summary}
\end{table}

For a like-for-like comparison, identical kinematic selections are applied to the partons and to the muons from the $Z'$ decay.
For the samples in which the $Z'$ boson is simulated at matrix-element level in \mg, namely \mgzphwshower\ and \mgzp, the two final-state partons from the hard process are used to define the kinematic observables. 
In contrast, for the samples where the $Z'$ boson is produced via the \hw shower, the identified partner parton and the recoiling hard parton are used.
Each parton is required to satisfy $p_T>30$~GeV and $|\eta|<2.5$, and each muon is required to satisfy $|\eta|<2.5$.

For samples in which the $Z'$ boson is generated using \mg, the kinematics receive sizeable contributions from ISR-like topologies and qqF processes, which are not present in the pure shower-induced radiation picture.
We note that ISR emission was also tested in \hw, however, in this case, the radiated $Z'$ boson is typically emitted close to the beam direction, such that the resulting muons fall outside the detector acceptance.
By contrast, ISR-like contributions present at the ME level can produce hard, central $Z'$ bosons.

To facilitate a meaningful comparison and to focus on the logarithmically enhanced regime relevant for parton shower radiation, the momentum-fraction distributions are therefore restricted to the collinear region defined by $\Delta R(q, Z')<1$.
This selection isolates the phase space where an interpretation in terms of a shower splitting is most appropriate.

After including the SM parton shower in \hw, the resulting predictions remain in good agreement with those reported in Ref.~\cite{herwig_bsm_ps}, in particular with the \mgzp\ and \hwonly\ configurations. 
In the $\Delta R(q, Z')$ distribution, \hw reproduces the FO prediction well in the collinear region for all configurations considered, as expected from the logarithmic structure encoded in the shower evolution. 
In contrast, at large $\Delta R(q,Z')$, the FO calculation provides the more appropriate description, since the parton shower framework is designed to capture soft and collinear radiation rather than wide-angle configurations.
A distinct feature is observed for \hwfull: as $\Delta R(q,Z') \to 0$, the yield increases and exceeds the corresponding \mg result.
This enhancement originates from additional $Z'$ emissions generated throughout the full shower evolution, which are absent in the pure matrix-element description. 
The behaviour therefore reflects genuine higher-order collinear contributions beyond the naive FO picture and motivates further dedicated study.

For the $z$ distribution, we observe an even closer match compared to the results of the previous paper, indicating that our restriction of the phase space to the collinear region is working as intended.
This level of agreement, however, gradually deteriorates for larger $Z'$ masses, as non-FSR contributions, which cannot be captured by a simple FSR-like picture, become increasingly important at higher mass.

The parton-level distributions shown above provide a first validation step and offer a clear picture of the underlying kinematics.
Building on this, to understand how these features manifest at the analysis-object level, in terms of reconstructed jets and leptons, we define observables analogous to those at the parton level and examine how well the corresponding trends are preserved.
Specifically, we map the two key variables, the angular separation and the momentum fraction of the quark-$Z'$ system, onto the angular separation between the jet and the $Z'$ candidate and the muon energy fraction inside the jet, the latter being an ingredient commonly used for jet identification in CMS~\cite{CMS-jet_id,CMS-pileup_mitigation}.
We consider jets with $p_T>30$~GeV that satisfy $\Delta R<0.4$ with respect to the partner quark, and we construct the $Z'$ candidate from the leading opposite-sign muon pair.

\begin{figure}
    \centering
    \includegraphics[width=0.49\linewidth]{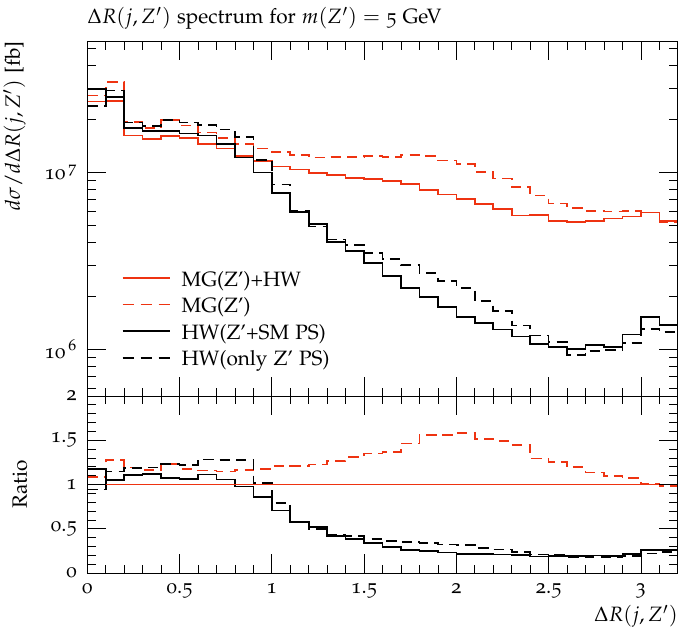}
    \includegraphics[width=0.49\linewidth]{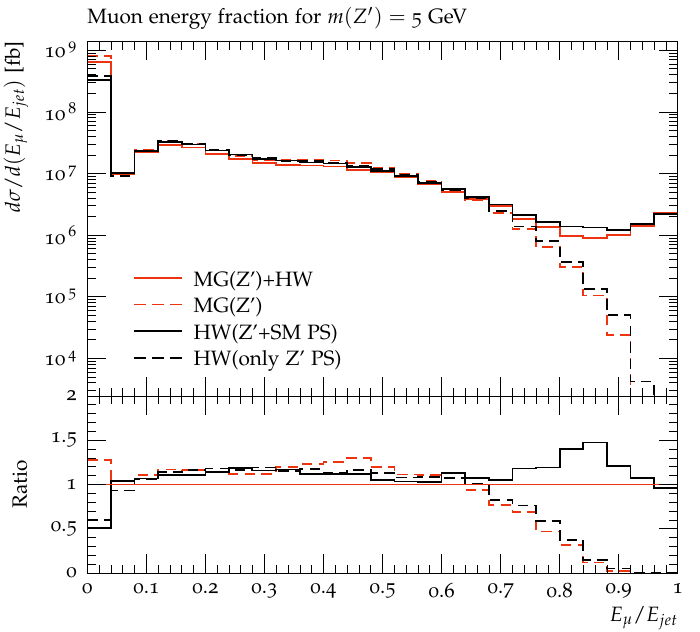}
    \includegraphics[width=0.49\linewidth]{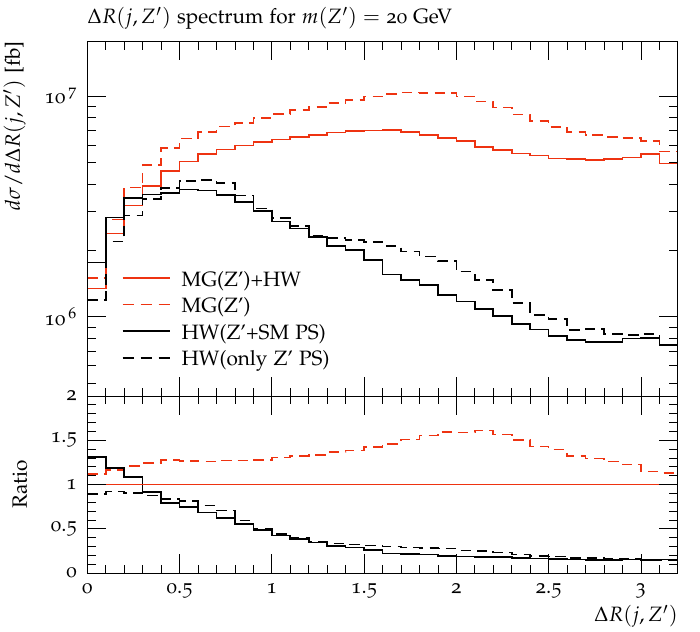}
    \includegraphics[width=0.49\linewidth]{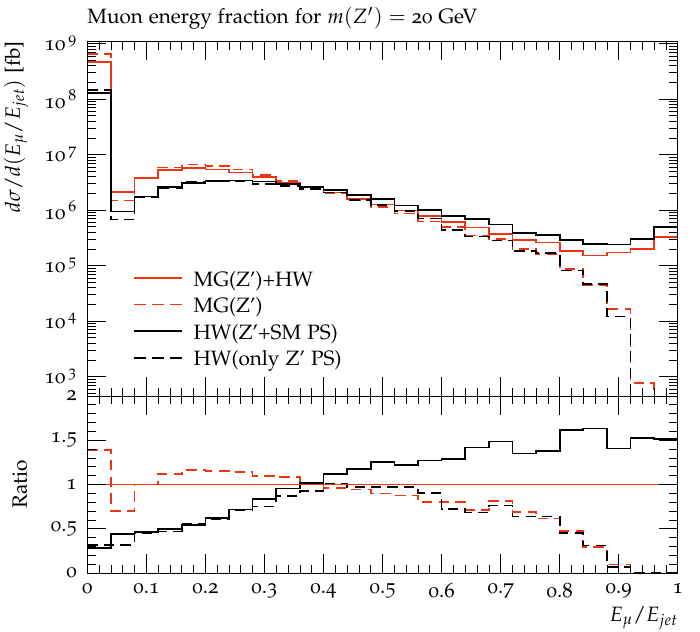}
    \includegraphics[width=0.49\linewidth]{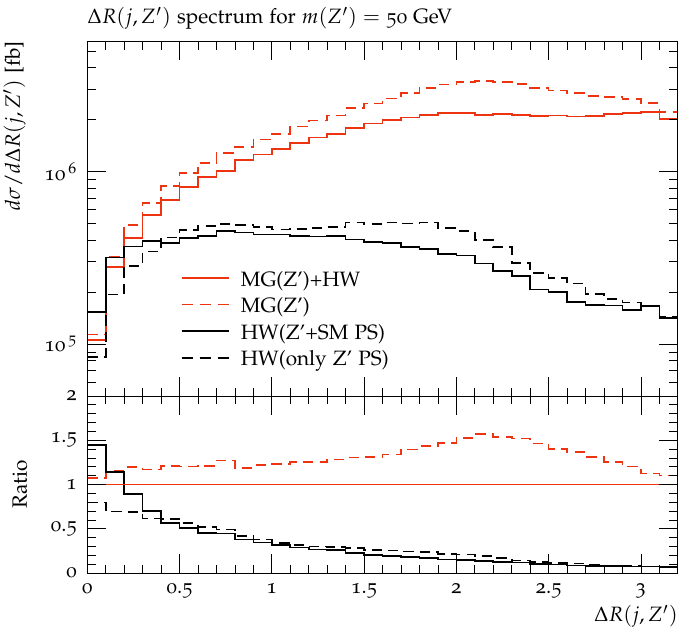}
    \includegraphics[width=0.49\linewidth]{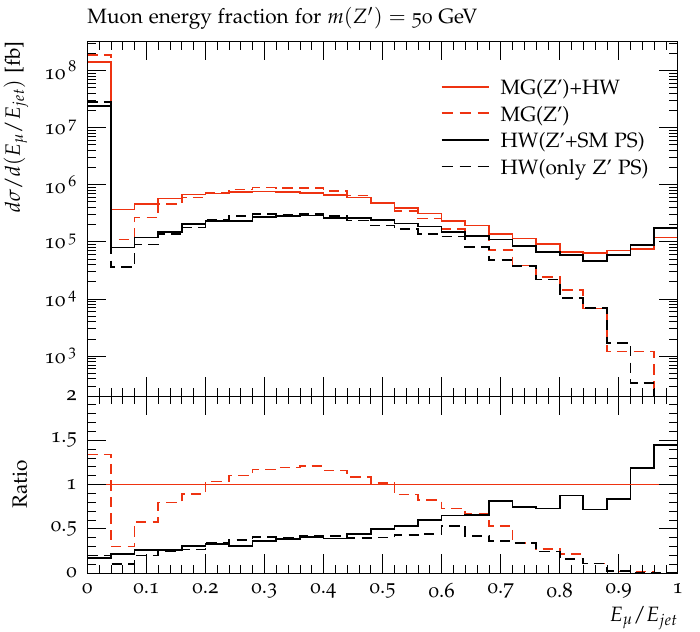}
    \caption{Analysis-level observables corresponding to the parton-level variables. The top, middle, and bottom panels show results for $M_{Z'}=5$, 20, and 50~GeV, respectively. Left: $\Delta R$ between the reconstructed jet and the di-muon system forming the $Z'$ candidate. Right: muon energy fraction inside the jet.}
    \label{fig: jet validation}
\end{figure}

Figure~\ref{fig: jet validation} presents the spectra of these two variables.
Although Fig.~\ref{fig: parton validation} shows distributions for the partons directly involved in the $Z'$ emission, whereas the present figure is constructed from jets and leptons after the full shower evolution, the two sets of plots can be interpreted coherently once object-level effects are taken into account.
For the $\Delta R$ distribution, not only is the overall shape similar, but the plot also captures the expected migration towards smaller $\Delta R$ in configurations where the partner parton and the $Z'$ are highly collimated.
In such cases, the muons from the $Z'$ decay can be clustered into the same jet, pulling the jet-$Z'$ angular separation closer to zero.

The muon energy fraction serves as an indirect proxy for the fraction of the $Z'$ energy carried inside the jet, and therefore roughly corresponds to $1-z$ in the parton-level $z$ spectrum, in particular in the limit where the parton and the $Z'$ are fully collimated.
An important difference is that, in the muon energy-fraction observable, muons that fall outside the jet do not contribute; consequently, such events accumulate at the endpoint.
At the other endpoint, as the muon energy fraction approaches unity, the distribution follows the $z\to 0$ behaviour of the parton-level spectrum in the absence of the SM shower, while after including the SM shower, the cross section no longer vanishes at the endpoint.
This reflects the fact that the partner parton can be softened by the SM shower and may partially leak outside the jet cone.

\begin{figure}
    \centering
    \includegraphics[width=0.49\linewidth]{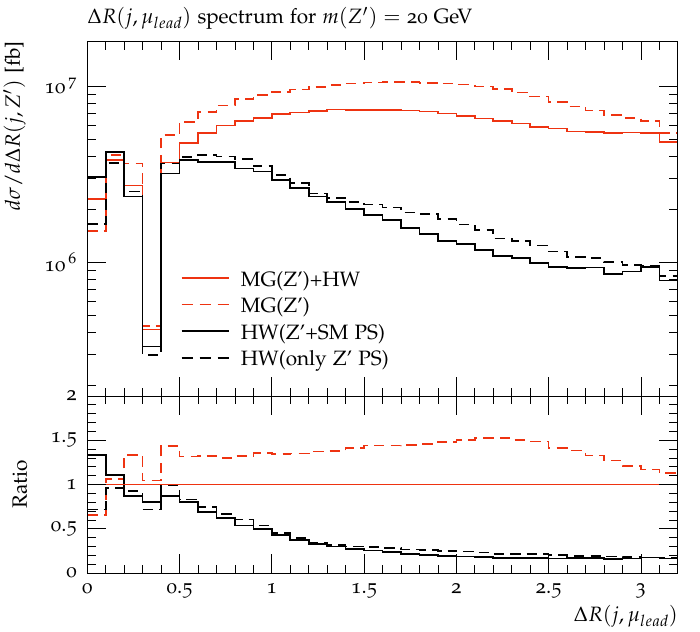}
    \includegraphics[width=0.49\linewidth]{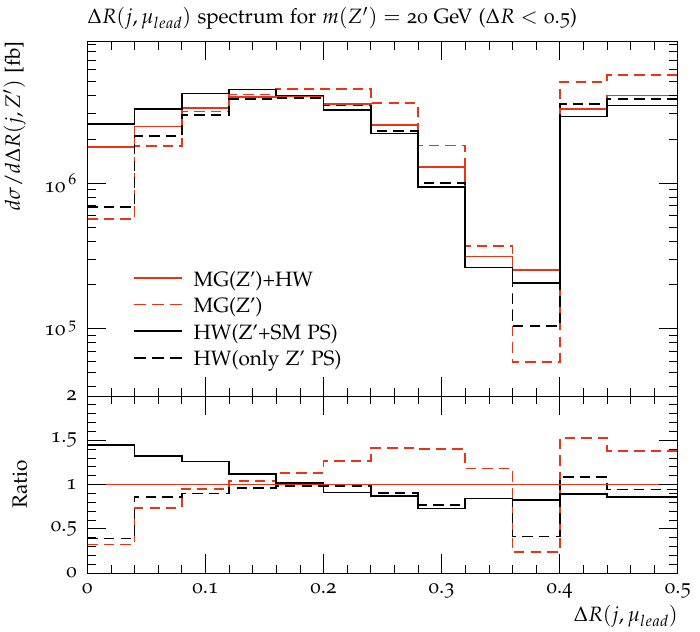}
    \includegraphics[width=0.49\linewidth]{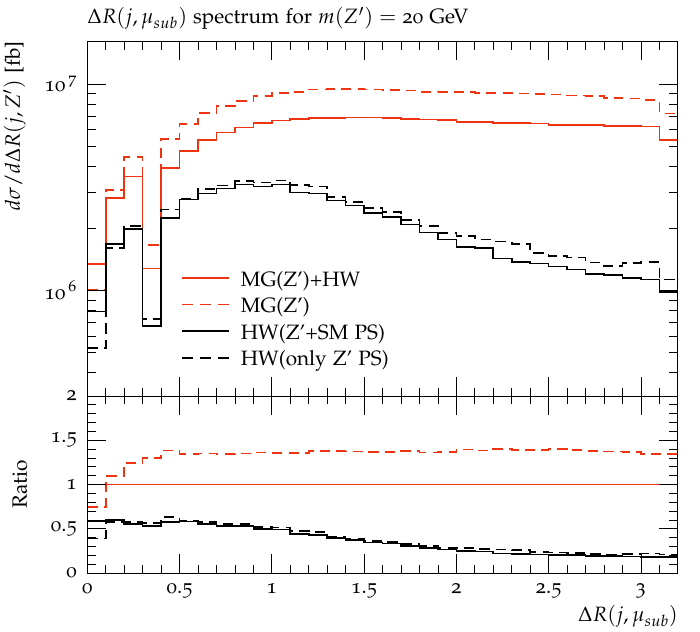}
    \includegraphics[width=0.49\linewidth]{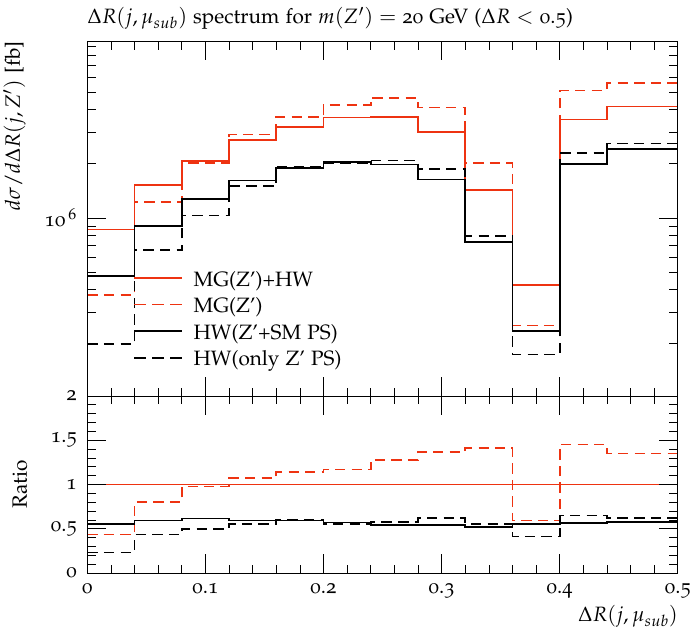}
    \caption{$\Delta R(\mu,\text{jet})$ distributions for $M_{Z'}=20$~GeV. The top (bottom) panels show the separation between the jet and the leading (subleading) muon. The left column covers the full range, $\Delta R<3.2$, while the right column provides a zoomed view of the near-jet region, $\Delta R<0.5$.}
    \label{fig: muon angle}
\end{figure}

Finally, Fig.~\ref{fig: muon angle} shows the angular separation, $\Delta R$, between the jet and the leading and subleading muons for the case $M_{Z'}=20$~GeV.
Since the muons from the $Z'$ decay are typically hard, the jet is often reconstructed around them, and therefore the events are more concentrated in the small-$\Delta R$ region, $\Delta R<0.4$.
A pronounced dip appears around $\Delta R\sim 0.4$; to elucidate its origin, we also show, on the right, a zoomed view of the jet-muon separation spectra restricted to $\Delta R<0.5$.
We find that the jet-muon separation falls smoothly as $\Delta R$ approaches 0.4 and then rises again beyond 0.4, which is a consequence of using the anti-$k_t$ (AK) jet cluster algorithm~\cite{AK_algorithm} with radius parameter $R=0.4$ for jet reconstruction.
Unlike an idealised cone-jet definition, which enforces a strict $\Delta R$ cut on the angle between the jet axis and the jet constituents, the anti-$k_t$ algorithm enforces a cut instead on the opening angle between pairs of constituents. 
When a constituent with a finite amount of energy near the edge of the jet is merged into it, the jet axis is recalculated and gets pulled towards that constituent. 
Energy flow, and in this case muon constituents above a given energy cut, are therefore suppressed as $\Delta R$ approaches the jet radius, in this case 0.4.
Therefore, in the following results, we adopt a conservative requirement $\Delta R(\mu,\text{jet})<0.3$ in order to focus on muons that are robustly associated with the jet and have not influenced its direction unduly.

\section{Results and Discussions} 
\label{sec:result}

We now quantify the experimental sensitivity to the $Z'$ radiation signatures considered in this study.
To this end, we evaluate the expected significance and derive projected exclusion limits as functions of the $Z'$ mass and coupling strength within the minimal $U(1)_{B-L}$ framework.
Given the exploratory nature of this analysis, we adopt a simplified statistical treatment based on a smooth analytic background parametrisation and the Asimov-data approach.

Signal events are selected by requiring at least one hard jet with $p_T>30$~GeV and $|\eta|<2.4$ that contains an oppositely charged muon pair.
For the standard single-muon trigger scenario, the muons are required to satisfy $p_T>52$~GeV for the leading muon and $p_T>5$~GeV for the subleading muon, where the high leading-muon threshold follows the CMS single-muon trigger requirement of approximately 50~GeV.
For the scouting-trigger scenario, both muons are required to satisfy $p_T>5$~GeV~\cite{cms-scouting-parking-run1,cms-scouting-parking-run2,EXO-19-018,EXO-21-005}.
In all cases, $|\eta|<2.4$ is imposed for each muon, and $\Delta R(j,\mu)<0.3$ is required to ensure that the muons are geometrically contained within the jet.
Throughout the analysis, conservative selections are adopted, both in the pseudorapidity acceptance and in the angular-separation requirements, in order to avoid overestimating the $Z'$ contribution inside jets, even though the upgraded CMS detector is expected to extend the muon coverage to $|\eta|<2.5$.

The background is modelled through an analytic fit to the di-muon spectrum of the simulated QCD- and top-induced samples.
The distribution exhibits a smoothly falling behaviour over the mass range of interest, described by a power-law function multiplied by a smooth turn-on factor accounting for the trigger $p_T$ threshold.
Accordingly, we model the background spectrum as
\[
f(m_{\mu\mu}) = A\, m_{\mu\mu}^{-n} \times \frac{1}{2}\left[1 + \mathrm{erf}\!\left(\frac{m_{\mu\mu}-\bar{m}}{\sqrt{2}\,\sigma}\right)\right],
\]
where $\mathrm{erf}(\cdot)$ denotes the error function with mean $\bar{m}$ and standard deviation $\sigma$.

To ensure numerical stability in the high-mass region, where the background statistics are limited, the fit is performed in logarithmic space.
Specifically, a $\chi^2$ function is constructed using the logarithm of the binned background yields, with the corresponding statistical uncertainties propagated to the logarithmic scale.
The fit ranges are chosen to avoid backgrounds from $B$-hadron decays and on-shell $Z$ boson production, and to exclude narrow regions around the $\Upsilon$ resonances, i.e. $m_{\mu\mu}\in[4,\,8]$ or $[12,\,60]$~GeV, corresponding to our search region with the resonance vetoes applied.

\begin{figure}
    \centering
    \includegraphics[width=0.49\linewidth]{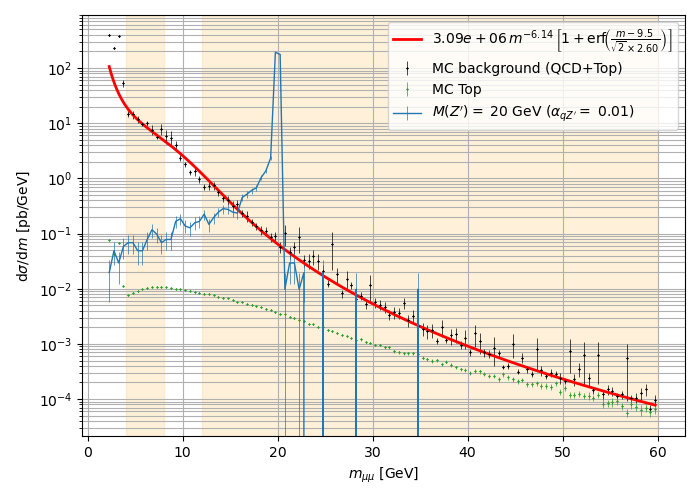}
    \includegraphics[width=0.49\linewidth]{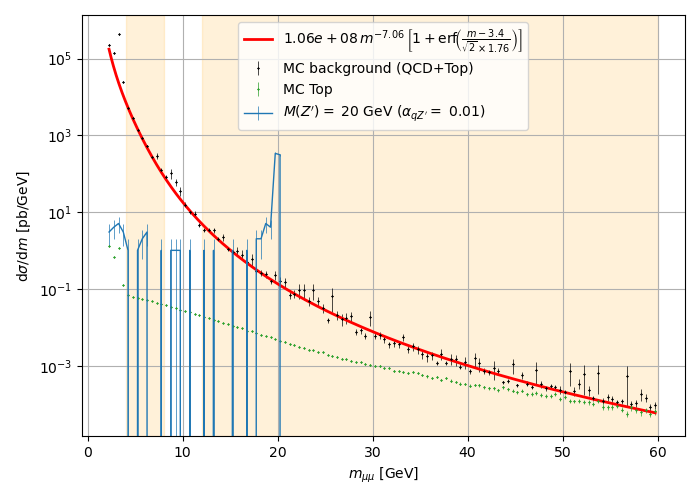}
    \caption{Di-muon invariant-mass distributions for the combined QCD and top backgrounds, together with the results of the analytic background fit.
    The top-induced background contribution and the signal distribution for $M_{Z'} = 20$~GeV with $\alpha_{qZ'} = 0.01$ are also shown for reference.
    Left: background spectrum obtained using the standard muon trigger at 13~TeV.
    Right: corresponding spectrum obtained using the scouting trigger strategy.
    The red curves indicate the fitted background functions, while the shaded regions denote the mass intervals used for the fit.
    The fit is performed in logarithmic space to ensure numerical stability in the high-mass region where the statistics are limited.}
    \label{fig: bkg fit}
\end{figure}

The resulting background distributions and the corresponding fits are shown in Fig.~\ref{fig: bkg fit} with an example signal distribution for $M_{Z'} = 20$~GeV and $\alpha_{qZ'} = 0.01$, where $\alpha_{qZ'} = g_{qZ'}^2/4\pi$. In both trigger configurations, the fitted function provides a good description of the smoothly falling background. For the nominal trigger, the error-function component exhibits a mean of $\bar{m}\simeq 8.4$~GeV with a relatively large width $\sigma\simeq 2.19$~GeV, reflecting the trigger turn-on behaviour. By contrast, in the scouting-trigger case, the fitted mean lies well below the minimum mass included in the fit, indicating that trigger turn-on effects are negligible over the fitted range.

The statistical significance of the $Z'$ radiation signal is evaluated using a counting experiment in a narrow invariant-mass window centred on the signal hypothesis.
For each $Z'$ mass point, the signal yield is obtained by integrating the simulated signal spectrum within a window defined as $\pm5\%$ around $M_{Z'}$, while the expected background yield is estimated by integrating the fitted background function over the same interval.
This window choice is intended to approximate the di-muon mass resolution inferred from the CMS muon transverse-momentum resolution~\cite{CMS_muon_performance}.

The expected significance is computed using the Asimov approximation,
\[
Z_{\mathrm{A}} = \sqrt{2\left[(S+B)\ln\left(1+\frac{S}{B}\right)-S\right]},
\]
where $S$ and $B$ denote the expected signal and background yields, respectively.
This definition provides an estimate of the median significance in the asymptotic limit and is therefore well suited to the first-look nature of this study~\cite{stat-asymptotic}.

\begin{figure}[t]
    \centering
    \includegraphics[width=0.49\textwidth]{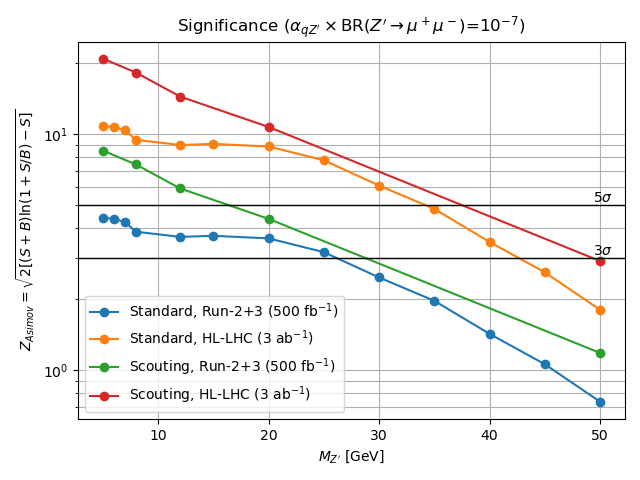}
    \includegraphics[width=0.49\linewidth]{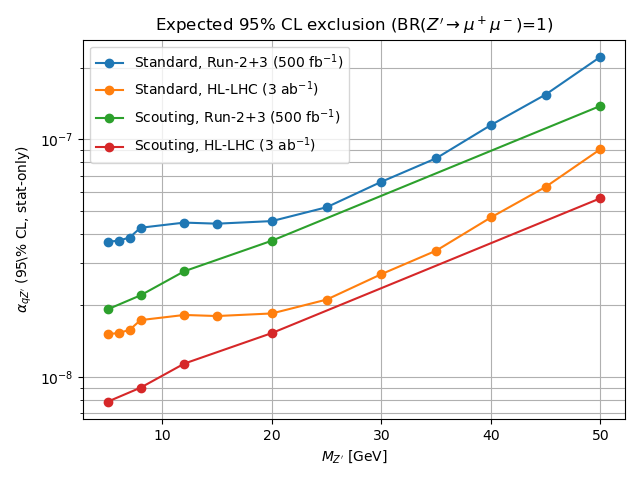}
    \caption{
    Expected Asimov significance for $Z'$ radiation inside jets as a function of $M_{Z'}$, assuming $\alpha_{qZ'} \times \mathrm{BR}(Z'\to\mu^+\mu^-)=10^{-7}$ (left), and the corresponding expected $95\%$ confidence-level exclusion limits on the $Z'$ coupling strength as a function of $M_{Z'}$ (right).
    The significance is evaluated in a fixed mass window around each hypothesis using the Asimov definition and includes statistical uncertainties only.
    Results are shown for the nominal muon-trigger and scouting-trigger strategies for Run-2+3 luminosity (500~fb$^{-1}$), and the corresponding HL-LHC projections (3~ab$^{-1}$) are shown as dashed curves with matching colours.
    }
    \label{fig:significance}
\end{figure}

The significance is evaluated separately for different trigger strategies and integrated luminosities.
Specifically, results are presented for the nominal muon trigger and for the scouting trigger configuration, each assuming integrated luminosities of 500~fb$^{-1}$, corresponding to the combined Run-2 and Run-3 dataset, and 3~ab$^{-1}$, representing the HL-LHC scenario.
This allows a direct comparison of the impact of trigger strategy and luminosity on the sensitivity to $Z'$ radiation inside jets.
Finally, Fig.~\ref{fig:significance} summarises the expected statistical sensitivity to $Z'$ radiation inside jets as a function of the $Z'$ mass hypothesis at $\alpha_{qZ'} \times \mathrm{BR}(Z'\to\mu^+\mu^-) = 10^{-7}$, and the corresponding expected $95\%$ confidence-level exclusion limits on the $Z'$ coupling strength as a function of $M_{Z'}$ under the $\mathrm{BR}(Z'\to\mu\mu)=1$ hypothesis.
As discussed in Sec.~\ref{subsec:signal}, in the regime of small $\alpha_{qZ'}$ considered in this study, the probability for $Z'$ emission in the parton shower scales linearly with $\alpha_{qZ'}$, while the observable di-muon signal rate is further proportional to $\mathrm{BR}(Z'\to\mu^+\mu^-)$.
The signal yield, and hence the significance, can therefore be straightforwardly rescaled for arbitrary values of the product $\alpha_{qZ'} \times \mathrm{BR}(Z'\to\mu^+\mu^-)$.

The trends indicate that the scouting strategy provides systematically higher sensitivity across the full mass range, consistent with its enhanced acceptance for low-$p_T$ muons.
The significance decreases with increasing $M_{Z'}$, reflecting the falling production rate and the reduced event yields at higher invariant masses in the relevant jet phase space.
In the low-mass region, $m_{\mu\mu}<20$~GeV, trigger-threshold effects are maximised, leading to a pronounced gap between the standard and scouting trigger strategies.
In this regime, the standard trigger suffers from a strong inefficiency for soft muons, while the scouting trigger retains high acceptance, resulting in an enhanced sensitivity.

At higher masses, the difference between the two trigger strategies increases again.
The reason is that the background composition becomes dominated by top-induced processes, which tend to produce muons with asymmetric transverse-momentum spectra, whereas the signal preferentially yields more symmetric muon $p_T$ configurations.
As a result, background events are more likely to satisfy the standard single-muon trigger requirements than the signal, thereby reintroducing a performance gap between the two trigger systems.

Building upon the significance estimates, we derive the expected exclusion limits on the $Z'$ coupling strength using the same statistical framework, shown in the right panel.
The limits are obtained under the background-only hypothesis by rescaling the signal yield until the Asimov significance reaches the one-sided $95\%$ confidence-level threshold, $Z_{\mathrm{A}}=1.64$.
The limits presented here are derived in a statistics-only approximation and should be interpreted as indicative of the intrinsic sensitivity of the proposed search strategy rather than as final experimental projections.

A detailed study of systematic uncertainties is beyond the scope of the present first-look analysis.
Nevertheless, if the observable of interest is defined as a ratio of cross sections,
\[
R(j\to Z') = \frac{\sigma(pp\to jj,\; j\to Z')}{\sigma(pp\to jj)},
\]
the dominant uncertainties associated with QCD scale variations and parton distribution functions are expected to be suppressed due to cancellation in the ratio.
The remaining theoretical uncertainties are anticipated to originate primarily from electroweak Sudakov logarithms associated with $Z'$ radiation and from the treatment of the bottom-quark mass scheme and fragmentation.
These effects are expected to be moderate and, in principle, amenable to systematic control in more refined future studies.

The observed trend follows the behaviour of the expected significance, with stronger exclusion limits obtained in the low-mass region where the $Z'$ radiation rate is enhanced.
At higher $Z'$ masses, the limits weaken due to the decreasing signal yield and reduced statistical power in the relevant phase space.
Across the full mass range, the scouting trigger strategy yields systematically stronger constraints than the nominal trigger.
These results serve as a first assessment of the discovery and exclusion potential of non-isolated di-muon signatures arising from $Z'$ radiation inside jets.
Even in this statistics-only framework, they provide a baseline for the study of parton shower-induced $Z'$ radiation and motivate more refined future studies incorporating systematic uncertainties and experimental effects.


\section{Summary and Conclusions}
\label{sec:summary}

This work has explored an experimentally under-investigated phase space for new physics searches, characterised by non-isolated di-muon signatures embedded inside jets.
Such signatures naturally arise when new neutral particles are radiated during parton shower evolution rather than being produced at the hard-scattering level.
Using the newly developed BSM parton shower framework in \hw, we generated fully showered event samples in which a light $Z'$ boson is emitted inside jets and subsequently decays into a collimated muon pair.
This provides a concrete demonstration that BSM parton shower radiation can act as a genuine signal mechanism, opening a complementary avenue for collider searches that is not covered by conventional isolated-lepton analyses.

Focusing on a minimal $U(1)_{B-L}$ benchmark as a representative example, we have shown that shower-induced $Z'$ radiation leads to distinctive kinematic features, most notably di-muons residing well inside jets.
Based on a smooth analytic description of the dominant QCD- and top-induced backgrounds, we have performed a first-look sensitivity study using Asimov-based significance estimates and expected exclusion limits.
Already with Run-2+3 luminosity and standard muon triggers, the analysis indicates that evidence-level sensitivity may be achievable for $Z'$ masses up to $\sim 25\,\mathrm{GeV}$ for $\alpha_{qZ'} \times \mathrm{BR}(Z' \to \mu^+\mu^-) = 10^{-7}$.
The reach is further enhanced when scouting-trigger strategies are employed, and projections to the HL-LHC suggest that searches across the $\mathcal{O}(10)\,\mathrm{GeV}$ mass range could become accessible.
Given the increasing jet activity at the HL-LHC and future hadron colliders, the importance of such non-isolated signatures is expected to grow.

This study establishes a baseline for a class of searches driven by BSM radiation in the parton shower and motivates several natural extensions.
These include interpretations in a broader set of models beyond the minimal $U(1)_{B-L}$ extension of the SM, such as two-Higgs-doublet models~\cite{2hdm-ufo,Darvishi:2017bhf, Darvishi:2019ltl, Darvishi:2020teg, Birch-Sykes:2020btk, Darvishi:2023nft}, and scenarios with enhanced couplings to third-generation quarks~\cite{pheno-RKZp+VectorNeutrino,pheno-RKZp}, as well as improvements from flavour tagging and vertex-based reconstruction of lepton pairs.
A dedicated treatment of systematic uncertainties, in particular electroweak Sudakov effects and heavy-flavour mass schemes, remains an important direction for future work.
Overall, our results demonstrate that BSM parton shower radiation provides a realistic and phenomenologically rich mechanism for uncovering new physics in regions that have so far remained largely unexplored.


\section*{Acknowledgments}

This research was supported in Korea by National Research Foundation grants RS-2024-00350406 and RS-2008-NR007227. This work has also received funding from the European Union's Horizon 2020 research and innovation programme as part of the Marie Skłodowska-Curie Innovative Training Network MCnetITN3 (grant agreement no. 722104). \textit{MRM} is supported by the UK Science and Technology Facilities Council (grant numbers ST/T001011/1 and ST/X000745/1), as is \textit{MHS} (grant numbers ST/T001038/1 and ST/X00077X/1).


\bibliographystyle{JHEP}
\bibliography{references}

\end{document}